\documentclass[preprint, 11pt, a4paper]{aastex631}
\usepackage[english]{babel}

\usepackage{amsmath}

\usepackage{graphicx}
\usepackage{natbib}

\newcommand{\kms}{~km~s$^{-1}$}
\newcommand{\obs}{_{\text{obs}}}

\newcommand{\E}{_{\text{E}}}
\newcommand{\glow}{_{\text{glow}}}
\newcommand{\lya}{Lyman-$\alpha$~{}}

\newcommand{\myvec}[1]{\ensuremath{\boldsymbol{{#1}}}}

\begin{document}
\title{WawHelioGlow: a model of the heliospheric backscatter glow. II. The helioglow buildup and the potential significance of the anisotropy in the solar EUV output}
\shorttitle{WawHelioGlowIIbuildup}
\shortauthors{Kubiak, Bzowski et al.}

\correspondingauthor{M. Bzowski}
\email{bzowski@cbk.waw.pl}

\author[0000-0002-5204-9645]{M. A. Kubiak}
\affil{Space Research Centre PAS (CBK PAN),\\
Bartycka 18A, 00-716 Warsaw, Poland}

\author[0000-0003-3957-2359]{M. Bzowski}
\affil{Space Research Centre PAS (CBK PAN),\\
Bartycka 18A, 00-716 Warsaw, Poland}

\author[0000-0002-6569-3800]{I. Kowalska-Leszczynska}
\affil{Space Research Centre PAS (CBK PAN),\\
Bartycka 18A, 00-716 Warsaw, Poland} 

\author[0000-0003-3484-2970]{M. Strumik}
\affil{Space Research Centre PAS (CBK PAN),\\
Bartycka 18A, 00-716 Warsaw, Poland} 

\begin{abstract}

The helioglow is a fluorescence of interstellar atoms inside the heliosphere, where they are excited by the solar EUV. Because the mean free path between collisions for the interstellar gas is comparable to the size of the heliosphere, the distribution function of this gas inside the heliosphere strongly varies in space and with time and is non-Maxwellian. Coupling between realistically modeled solar factors and the distribution function of interstellar neutral gas is accounted for in a helioglow model that we have developed. WawHelioGlow is presented in the accompanying Paper I. Here, we present the evolution of the gas density, solar illumination, helioglow source function, and other relevant parameters building up the helioglow signal for selected lines of sight observed at 1 au. We compare these elements for various phases of the solar cycle and we present the sensitivity of the results to heliolatitudinal anisotropy of the solar EUV output. We assume a realistic latitudinal anisotropy of the solar wind flux using results from analysis of interplanetary scintillations. We compare the simulated helioglow with with selected maps observed by the SOHO/SWAN instrument. We  demonstrate that WawHelioGlow is able to reproduce fundamental features of the sky distribution of the helioglow. For some phases of the solar cycle, the model with an anisotropy of the solar EUV output better reproduces the observations, while for other phases no EUV anisotropy is needed. In all simulated cases, the solar wind anisotropy following insight from interplanetary scintillation measurements is present.

\end{abstract}

\section{Introduction}
\label{sec:intro2}
\noindent
The heliospheric backscatter glow is the fluorescence of interstellar neutral (ISN) gas inside the heliosphere, resonantly excited by the solar EUV radiation. Discovery of a diffuse EUV background \citep{bertaux_blamont:71, thomas_krassa:71} was, in fact, crucial to the discovery of the heliosphere \citep{blum_fahr:70b}. Initially, investigations of the helioglow of ISN H and He were mostly focused to determine the flow vector and the temperature of ISN gas. Discovering that the temperatures and flow vectors of ISN H and He seem to differ from each other led to the conclusion that there must be a region of strong interaction between the plasma and neutral gas which results in a processing of ISN H but not of ISN He \citep{lallement_bertaux:90a, lallement_etal:93a, lallement:96}. Very soon it was realized that the \lya helioglow bears imprints of the heliolatitudinal structure of the solar wind, which can be retrieved by appropriate analysis of the helioglow distribution in the sky \citep{witt_etal:79,witt_etal:81, lallement_etal:85a, lallement_etal:86,lallement_etal:95b, lallement_etal:10a, bzowski_etal:03a, summanen_etal:97, katushkina_etal:13a, katushkina_etal:19a, koutroumpa_etal:19a, pryor_etal:96, pryor_etal:98a}. However, in all these papers, with the notable exceptions of the studies by Pryor et al., it was assumed that the solar wind varies with heliolatitude, but the solar EUV output does not. 

 An alternative method to study the solar wind structure is analysis of interplanetary scintillations (IPS). The method of interplanetary scintillation solar wind tomography relies on observationally discovered correlation between the level of the solar wind electron density fluctuations and the solar wind speed \citep{hewish_etal:64a, coles_maagoe:72a, coles_rickett:76a, kojima:79a, asai_etal:98a}. The density fluctuations are observed as scintillations in the time series of compact radio sources distributed in the sky nearby the Sun. This method is independent of both the ISN gas distribution inside the heliosphere and the assumptions on the latitudinal behavior of the solar EUV output; so far it has only been able to provide the solar wind speed at various heliolatitudes, but not the density. 

Based on solar wind profiles obtained from tomographic analysis of IPS observations \citep{jackson_etal:98a, tokumaru_etal:10a}, it was possible to study the evolution of the solar wind velocity structure over the two past solar cycles \citep{sokol_etal:13a, sokol_etal:20a}. However, detailed conclusions from these studies are sometimes at odds with those obtained from the helioglow analysis \citep{katushkina_etal:13a, katushkina_etal:19a}. 

Here, we present results of application of the WawHelioGlow code, extensively presented in Kubiak et al. 2021 (Paper I \footnote{M. A. Kubiak et al., WawHelioGlow: a model of the heliospheric backscatter glow. I. Model definition}) to the calculation of the helioglow of H and He, observed at 1 au from the Sun. We discuss the behavior of the factors making up the helioglow along selected lines of sight. We point out important differences depending on the geometry and solar cycle phase of the observations on one hand, and on the species (H or He) on the other hand. We pay special attention to the expected signatures of a heliolatitudinal anisotropy of the solar EUV output and we qualitatively demonstrate that including this effects may improve the agreement of the simulated distribution of the helioglow in the sky with measurements. In addition to the \lya glow, we briefly present the similarities and differences in the formation of the helioglow of helium. In this latter case, the hypothetical EUV anisotropy is expected to be the principal source of heliolatitude structure of the glow because photoionization is the main source of ionization losses. However, this effect is expected to be well visible only relatively close to the Sun, which may be a challenge for numerical codes. We demonstrate that WawHelioGlow is able to address these challenges without difficulty. 

The paper should be read with Paper I at hand since we frequently refer to its formulae and figures. 

\section{Calculations}
\label{sec:calculations}
\begin{figure}[ht!]
\centering
\includegraphics[width=0.8\textwidth]{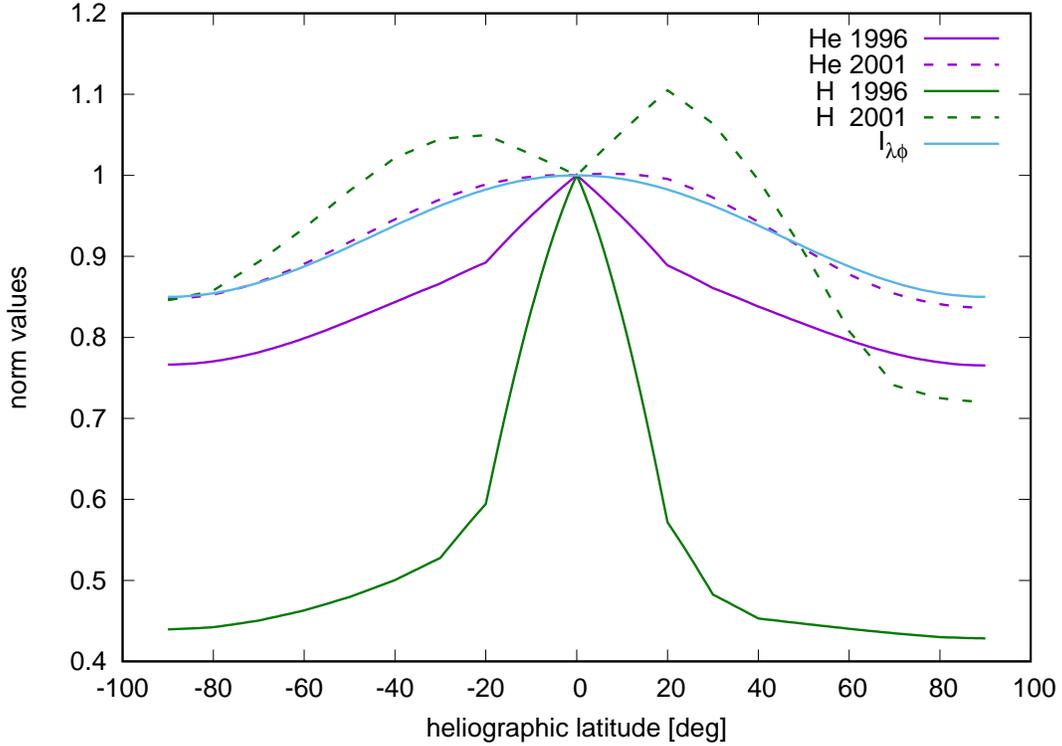}
\caption{Heliolatitudinal profiles of the total ionization rates of H and He (green and purple, respectively) for the epochs of simulations: 1996.43 and 2001.43 (solid and broken lines, respectively), and the anisotropy of the solar illumination of the gas (blue; see the function $I_{\lambda \phi}(\phi)$, Equation 29 in Paper I). All presented profiles are normalized to the 0\degr{} heliolatitude. The total ionization rates for this latitude for H were $8.235\times 10^{-7}\,\text{s}^{-1}$ for 1996 and $5.516\times 10^{-7}\,\text{s}^{-1}$ for 2001. For He, they are, respectively, $1.292\times 10^{-7}\,\text{s}^{-1}$ and $1.718\times 10^{-7}\,\text{s}^{-1}$.} 
\label{fig:helioLatProfiles}
\end{figure}

\noindent
The WawHelioGlow code is based on the (n)WTPM model of interstellar neutral (ISN) gas inside the heliosphere \citep{tarnopolski_bzowski:09, sokol_etal:15a}. The code for H is employed with the radiation pressure taken from \citet{IKL:20a}, which uses the composite solar \lya flux \citep{machol_etal:19a}. The ionization model is adapted from \citet{sokol_etal:20a}. Additionally, electron impact ionization is included following \citet{bzowski_etal:12b} based on the OMNI2 solar wind data \citep{king_papitashvili:05}.  
The EUV-related factors, i.e., the illumination of ISN gas by the Sun, the photoionization rate, and the radiation pressure assume that the solar flux is anisotropic in heliolatitude (see Figure \ref{fig:helioLatProfiles} and Equation 29 in Paper 1). 

The calculations for hydrogen are performed assuming that the population of ISN atoms responsible for the helioglow comprises two sub-populations: the primary and the secondary, with the parameters identical to those listed in Table~1 in \citet{IKL:18b}. The simulations were performed for these two populations separately. Because of the adoption of the single-scattering, optically thin approximation, the results of calculations for individual populations can be arithmetically co-added to yield a full helioglow signal. The results from the two-population model are compared with a one-population approach with carefully selected parameters to best match the two-population case. For helium, the one population model was used.

If the solar factors, i.e., ionization, radiation pressure, and illumination featured spherical symmetry, the ISN gas distribution would show an axial symmetry around the upwind--downwind line. Since this latter line is inclined at an angle of $\sim 5\degr$ to the ecliptic plane, and we consider observations performed at 1~au from the Sun in the ecliptic plane, the expected distribution of the helioglow in the sky would show certain characteristic asymmetries. In our model, however, we do not assume a spherical symmetry of the solar factors, as illustrated in Figure~\ref{fig:helioLatProfiles}. 

In this paper, we used a model of the ionization factors based on available observations of the solar factors (solar wind and the EUV radiation). The total ionization rate is a sum of the rates of photoionization, charge exchange of ISN atoms with solar wind protons and, for He, additionally with solar wind alpha particles, and of electron impact. The charge exchange rate is based on the model of evolution of the solar wind speed and density based on IPS observations. This reaction is the dominant one for H and almost negligible for He. Photoionization is the second important factor for H and the dominant one for He. The latitudinal profile of the photoionization rates for H and He are adopted from \citet{sokol_etal:20a} as the in-ecliptic rates for a given time. Beyond the ecliptic the rates are modulated by the function given in Equation 29 in Paper I with the parameter $a$ equal either to 0.85, which results in the profile shown with the blue line in Figure \ref{fig:helioLatProfiles}, or to 1, which yields a spherically symmetric rate. Electron impact rate is the least intense for both H and He \citep{rucinski_etal:96a}, and its relative contribution to the total rate rapidly drops with the solar distance, becoming negligible outside a few au \citep{rucinski_fahr:91}. Its rate is proportional to the total electron density in the solar wind, equal to a sum of the proton and doubled alpha densities. Consequently, its latitudinal profile follows that of the solar wind density. Here, we neglect the difference in the electron-impact rates for the slow and fast wind conditions \citep[see, e.g.,][]{bzowski_etal:12b}. 

The latitudinal profiles of the total ionization rate may be complex in shape and vary with time. Example profiles for the solar minimum and maximum for the two discussed species are presented in Figure \ref{fig:helioLatProfiles}. As a result of the ionization rate anisotropies, the helioglow distribution in the sky features additional asymmetries.   

We consider the ionization model as a realistic one. In its construction, we deliberately did not use any insight from analysis of helioglow observations. With this, we are able to verify to what extent the helioglow is sensitive to anisotropies of the EUV.
We also investigate to what extent a helioglow model is capable of reproducing helioglow observations, if this model is based on first principles and only on available observations of relevant quantities that are not based on helioglow-related information.
This comparison is briefly presented in Section~\ref{sec:SWANcomparison}. Before that, however, we demonstrate the behavior of the factors building up the signal (Sections \ref{sec:SourceFunctionElements}--\ref{sec:illuminationFlattening}) along carefully selected lines of sight (Section \ref{sec:grid}).

\section{Buildup of the signal from hydrogen}
\label{sec:SimulationsH}
\noindent

\subsection{Selection of lines of sight and the simulation grid }
\label{sec:grid}
\begin{figure}[ht!]
\centering
\includegraphics[width=0.7\textwidth]{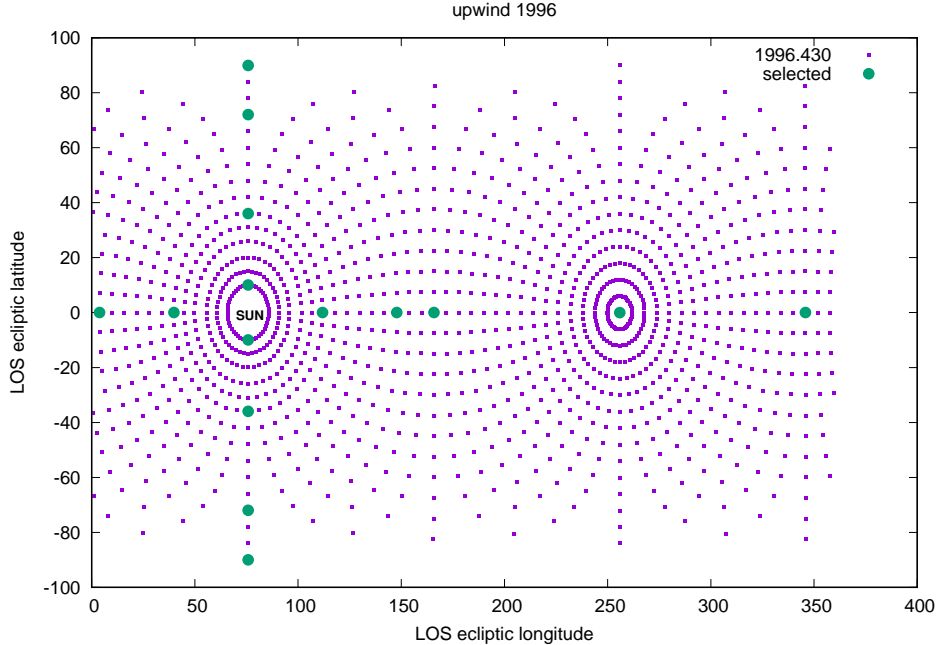}
\caption{The simulation grid for a map of the helioglow for an observer located at 1 au at longitude 255.85\degr{} (close to the upwind direction). The direction towards the Sun is thus towards longitude 75.85\degr. The directions of lines of sight are located at concentric rings (small purple rectangles) centered at 4\degr{} off the Sun's direction. The empty space is the exclusion zone around the Sun. The large green dots represent the selected lines of sight (see text), located at both sides of the Sun in the ecliptic plane and in the polar plane.} 
\label{fig:gridSelect}
\end{figure}

\noindent
Calculation of the intensity of the helioglow involves two spatial parameters: the observer location in space $\myvec{r}\obs$ and the direction into which the observer is looking $\myvec{\hat{l}}$ (Equation 3 in Paper I). The observer locations were chosen to facilitate referring the simulation results on one hand to the available maps of the helioglow obtained by the SWAN experiment, and on the other hand to the geometry of the planned experiment GLOWS. SWAN takes full-sky maps of the helioglow approximately once a day, with two exclusion zones: one centered at the Sun, and another one around the spacecraft, usually, in the anti-solar direction \citep{bertaux_etal:97a}. GLOWS scans a small circle with a radius of 75\degr, centered in the ecliptic plane at a longitude shifted by 4\degr{} relative to that of the Sun. The simulation grid (Figure~\ref{fig:gridSelect}) features concentric rings around the center of the planned GLOWS scanning circle. They cover the entire sky except a region within 36\degr{} off the Sun. The absolute ecliptic coordinates of the grid points  vary with the location of the observer. 

Out of this grid, we select several characteristic lines of sight, marked as large green dots in Figure~\ref{fig:gridSelect}. They include the antisolar line as well as lines at $\pm 36\degr, \pm 72\degr$, and $\pm 90\degr$ in the ecliptic plane and in the plane perpendicular to the ecliptic, going through the northernmost and southernmost points in the abovementioned rings. This makes a total of six lines of sight selected in each of the two perpendicular planes for each observer locations. 

In the following, we analyze elements building up the source function of the helioglow by placing the observer in the ecliptic plane approximately at the upwind, downwind, and crosswind location for observations performed on the days of the year for which SWAN maps are available. These days were chosen during the minimum (the epoch 1996) and the maximum of the solar activity (the epoch 2001). 

\begin{figure}[ht!]
\centering
\includegraphics[width=0.47\textwidth]{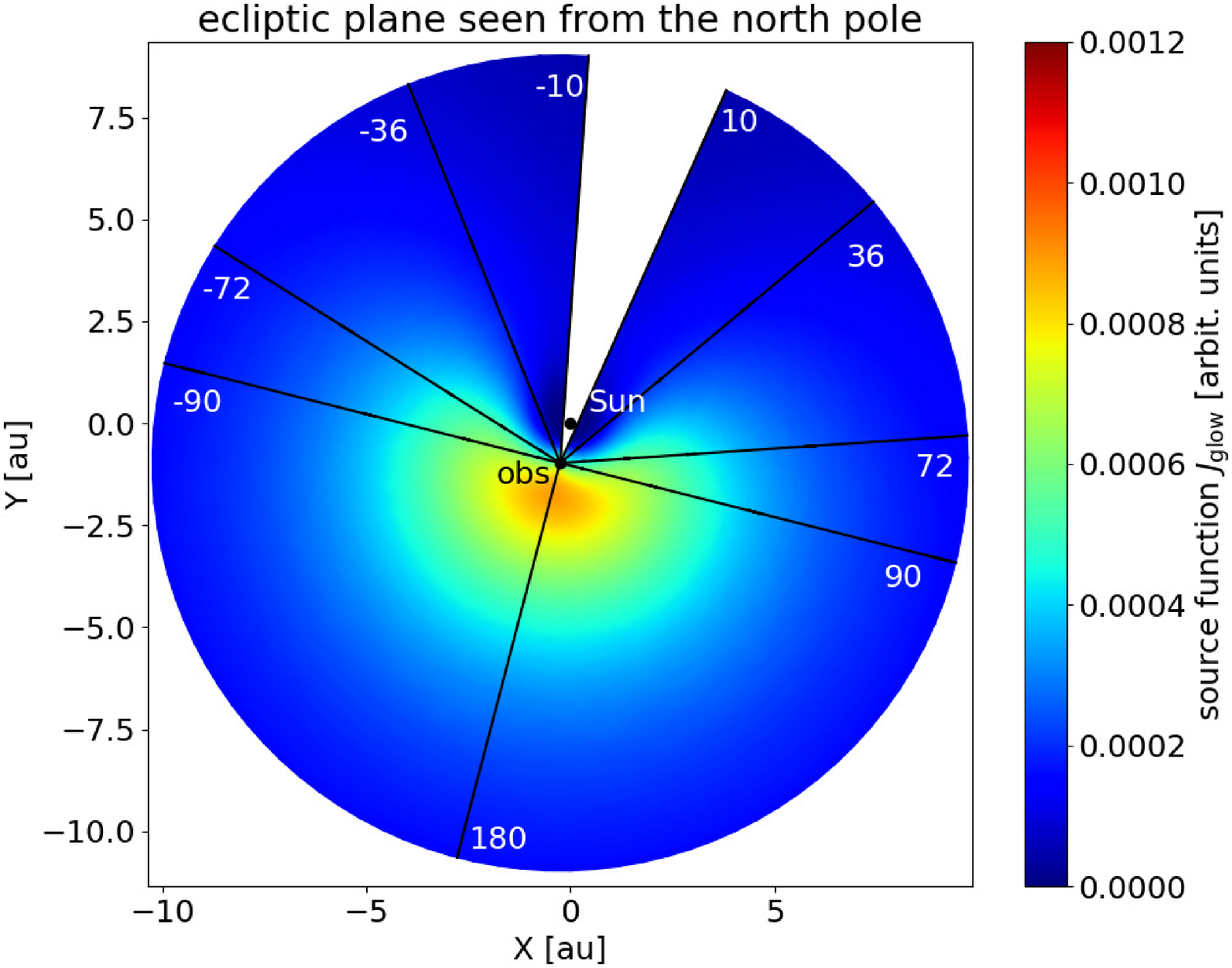}
\includegraphics[width=0.47\textwidth]{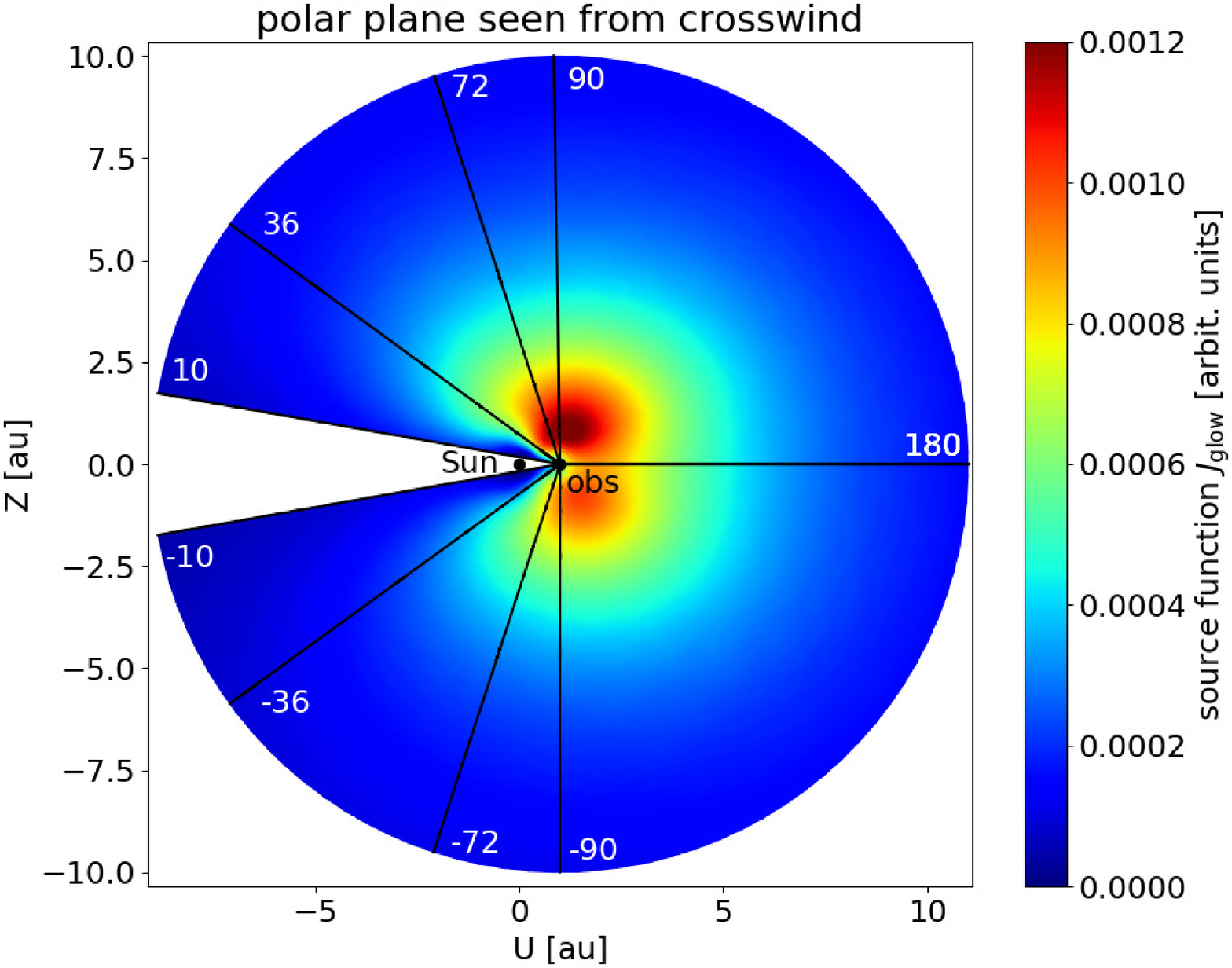}
\includegraphics[width=0.47\textwidth]{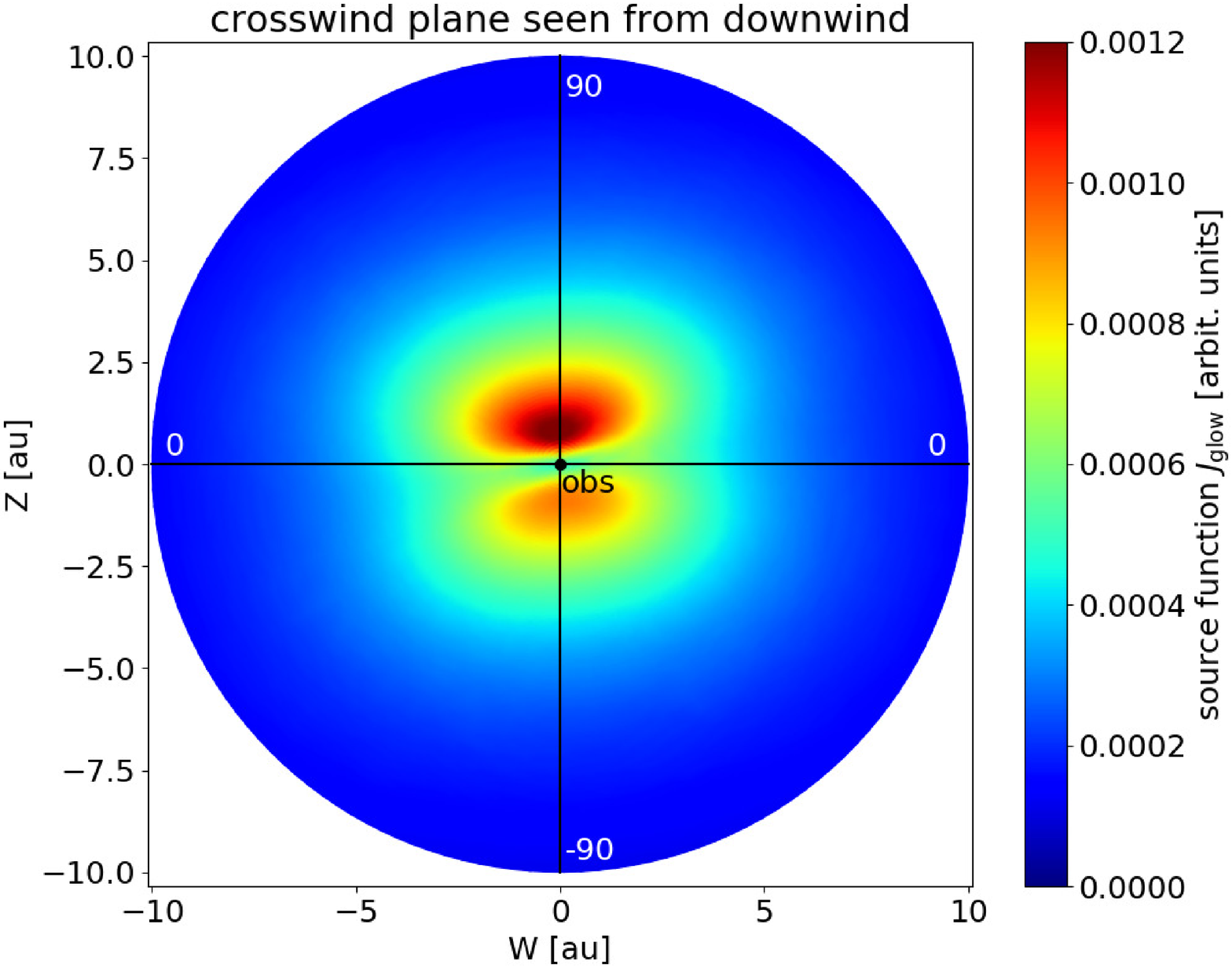}
\caption{Color-coded magnitudes of the source function of the hydrogen glow for the primary population in three selected planes centered at the observer. The reference system is Cartesian, with the Sun at the origin.
The upper left panel shows the ecliptic plane seen from the north ecliptic pole. The observer (marked with the black circle marked up ``obs'') is at the upwind direction at 1 au from the Sun, at the corner of the white $\pm 10\degr${} exclusion zone. 
In the upper right panel, a polar plane is seen from the crosswind direction in the ecliptic plane. The polar plane includes the inflow direction and the north-south axis, but in the projection presented in the upper right panel he is now to the right from the Sun. 
The lower panel presents a plane perpendicular to the upwind-downwind direction, seen from the downwind direction. The observer is located within this plane, which implies that the plane is offset by 1 au upwind from the Sun.
All three panels are in a Cartesian reference frame, with the axes scaled in au. The black lines represent selected lines of sight. The color scale is in arbitrary units. }  
	\label{fig:densPlane}
\end{figure}

To facilitate understanding the spatial distribution of these lines of sight and the helioglow source function, we present the distribution of the source function in three selected heliocentric planes (see Figure \ref{fig:densPlane}), color coded and with superimposed selected lines of sight. The source function features well-defined maxima in space (red and dark-red regions), where formation of the helioglow is the most intense. Isocontours of the source function reflect on one hand the geometry of the gas flow, and on the other hand, the imprints of the ionization losses in the ISN gas. 

Figure~\ref{fig:densPlane} clearly shows the symmetries and asymmetries of the system. In the ecliptic plane (upper left panel) there is a symmetry in the source function distribution relative to the solar direction. Since the observer is at the upwind line, the upwind line is for him antisolar. Consequently, the symmetry of the ISN H density relative to the upwind direction and the symmetry of the ionization losses and the solar illumination coincide.  

The upper-right panel in Figure \ref{fig:densPlane} shows a plane including the two ecliptic poles. 
A north-south asymmetry relative to the ecliptic plane is clearly visible: the region of the highest magnitude of the source function is larger in the northern hemisphere than in the southern. This is because of the $\sim 5\degr${} inclination of the flow direction of ISN gas to the ecliptic plane and to the solar equator, and because we use a 3D model of the ionization losses, which results in a different densities in the north and south ecliptic hemispheres for the selected time moment. As a result, a more abundant gas yields a larger source function for the helioglow, because the solar illumination is assumed to be north-south symmetric relative to the solar equator. 

A similar effect is visible in the lower panel of Figure~\ref{fig:densPlane}, which shows a plane perpendicular to the direction towards the Sun. This panel demonstrates that the density distribution is organized relative to the solar equator plane and not the ecliptic plane.

\subsection{Individual elements of the source function along selected lines of sight}
\label{sec:SourceFunctionElements}
\noindent
In this section, we analyze individual elements building up the source function $J\glow$ (Equation 2 in Paper I) and their relations depending on the observer location and the direction of the line of sight. 

\begin{figure}[ht!]
\centering
\includegraphics[width=0.97\textwidth]{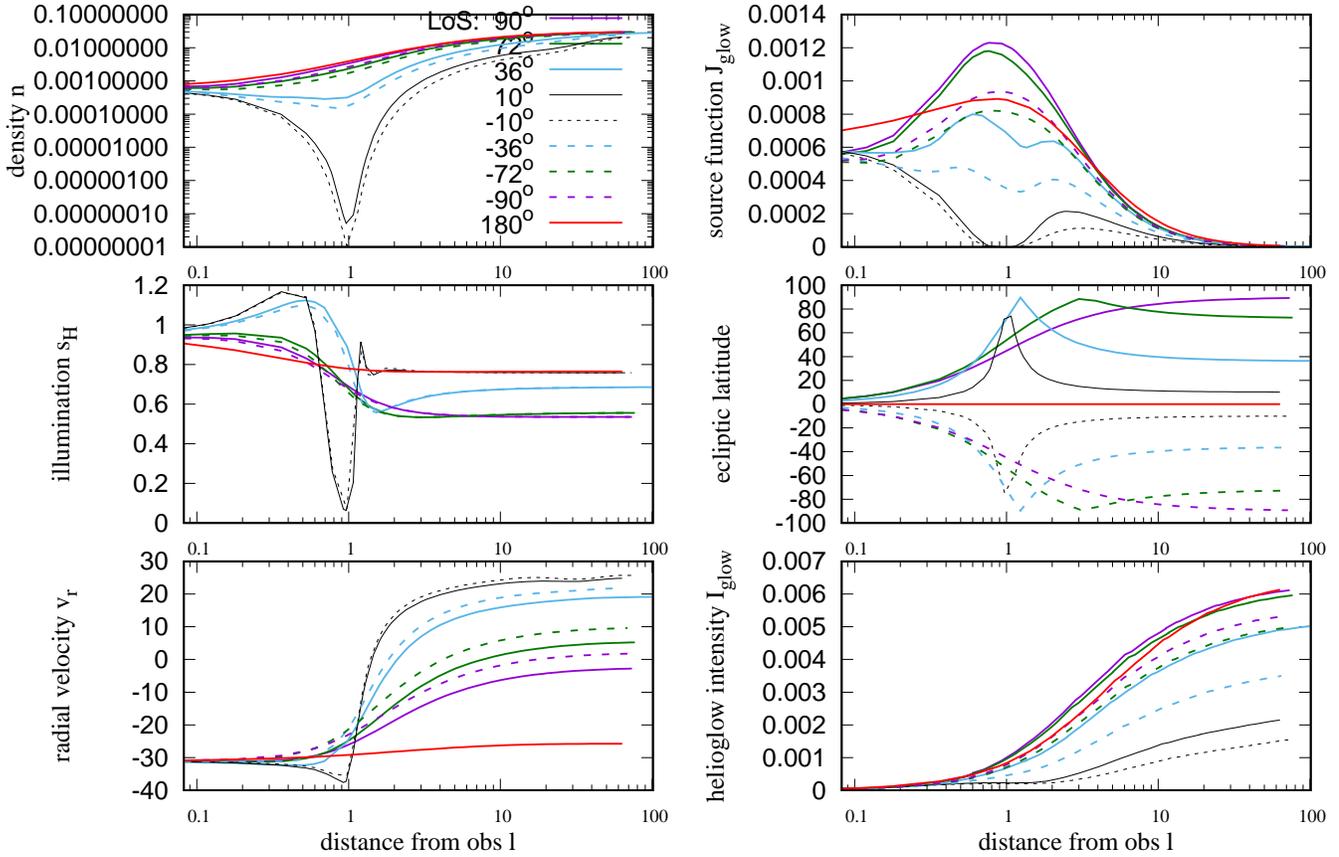}
\caption{Variation of the elements of the source function of the helioglow along the line of sight for the conditions of solar minimum: 
the density of ISN H ($n_\text{H}(l)$; upper left panel; cm$^{-3}$), 
the illumination $s_\text{H}$ (arbitrary units), 
the mean radial velocity of the gas relative to the Sun ($\langle v_r(l)\rangle$; lower left panel; \kms),
the source function ($J\glow(l)$, Equation 2; upper right panel; arbitrary units), 
the latitude of the line of sight at $l$ (middle right panel; degree), 
and the partial helioglow intensity ($I\glow(l)$, Equation 35; lower right panel; arbitrary units).
The equation numbers refer to Paper I. The units are identical for the given quantities shown in all figures.
Note that with our definitions, $s_\text{H}$ is equal by number to the radiation pressure compensation factor $\mu(l)$ (Equation 33), calculated for $\langle v_r(l)\rangle$.
The observer is located at the upwind position at 1 au in the ecliptic plane, the epoch (1996.43) corresponds to the minimum of the solar activity between solar cycles 22 and 23, the ISN H parameters correspond to the primary population. The horizontal axis measures the distance from the observer in au along the line of sight. The observer is scanning the polar plane to the north and to the south of the ecliptic plane (cf. Figure \ref{fig:gridSelect} and the upper right panel in Figure \ref{fig:densPlane}). The northern and southern lines of sight are given in solid and broken lines, respectively. The 180\degr{} line of sight is antisolar.}
\label{fig:upminPolespr}
\end{figure}
\begin{figure}[ht!]
\centering
\includegraphics[width=0.97\textwidth]{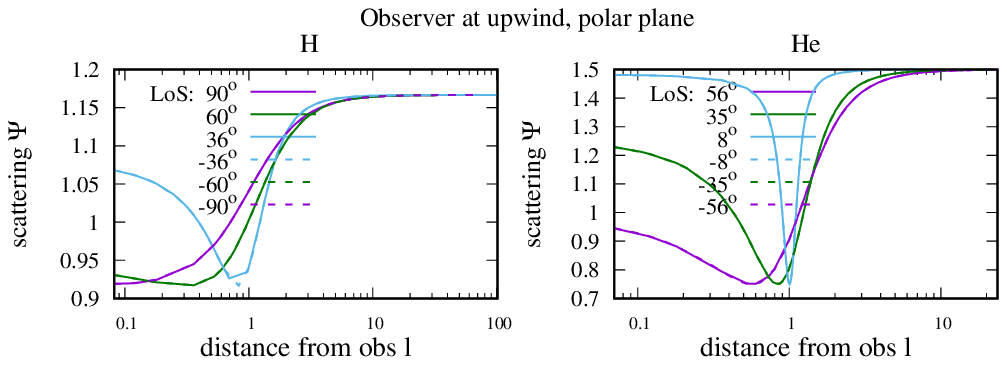}
\caption{Scattering phase function $\psi(l)$ along the line of sight for H  and He (left and right panels, respectively; Equations 8 and 9 in Paper I) for the upwind vantage point. For the other vantage points discussed in the paper, $\psi(l)$ are practically identical.}
\label{fig:HeupminPsi}
\end{figure}

Figure~\ref{fig:upminPolespr} presents the elements forming the signal along the selected lines of sight: the gas density, the solar illumination that excites this gas (which is proportional to the local radiation pressure factor $\mu$; see Section 2.6.1 in Paper I), the mean radial velocity of the gas relative to the Sun (Equation 12 in Paper I), the latitude along the line of sight, and the source function of the helioglow (Equation 2 in Paper I). The figure is completed by the partial helioglow intensity $I\glow(l)$ (Equation 35 in Paper I, first row, $\int\limits_{0}^L J\glow d l$, cumulative intensity from the observer to a given distance). The phase function $\psi_\text{H}$ for H (Equation 8 in Paper I) is presented in the left panel of Figure \ref{fig:HeupminPsi}. Note that for the lines of sight discussed in this paper, $\psi(l)$ depends almost solely on the elongation of the line of sight from the Sun (i.e., on the angular separation of the aiming point of the line of sight from the Sun's center) and, consequently, $\psi(l)$ is almost identical for the equivalent line of sight elongations for all vantage points discussed. The same is true for the helium lines of sight, discussed in Section \ref{sec:SimulationsHe}.

The phase function modifies the source function relatively little, since its (minimum, maximum) values are equal to (0.90, 1.17), as shown in Figure 2 in Paper I. The magnitude of the illumination of the gas drops with the square of solar distance, but additionally is modulated by the radial speed of the gas (cf. the middle and lower left panels in Figure~\ref{fig:upminPolespr}). In the case presented in this figure, the observer is upwind from the Sun and is looking downwind. Consequently, the speed of the gas varies from extreme negative close to the observer (the gas approaching the Sun) to extreme positive (the gas receding from the Sun far from the observer). For the secondary population, the qualitative behavior is similar, but the amplitude is lower by 5--10~\kms{} because the inflow speed of the secondary population is lower than that of the primary. The modulation of the illumination in this observation geometry is from $\sim 0.5$ to $\sim 1.1$, i.e., the variation is approximately by a factor of 2. This illustrates the need to take the actual form of the solar \lya line into account. 

The factor that features the largest amplitude along the line of sight is the density (upper left panel in Figure \ref{fig:upminPolespr}), which in the geometry used in this figure varies by more than two orders of magnitude. This causes the source function to significantly vary, featuring a maximum within $\sim 1$ au from the observer (depending on the offset of the line of sight from the Sun), going down by an order of magnitude at $\sim 10$ au. As illustrated in the lower right panel, approximately half of the magnitude of the helioglow in this viewing geometry is formed within 3---5 au from the observer, and there is a clear modulation of the source function as the line of sight is the nearest to the Sun, which in this viewing geometry is between $\sim 0.75$ au from the observer for the offset of 36\degr{} and $\sim 0.4$ au for an offset of 72\degr.

\begin{figure}[ht!]
\centering
\includegraphics[width=1.3\textwidth]{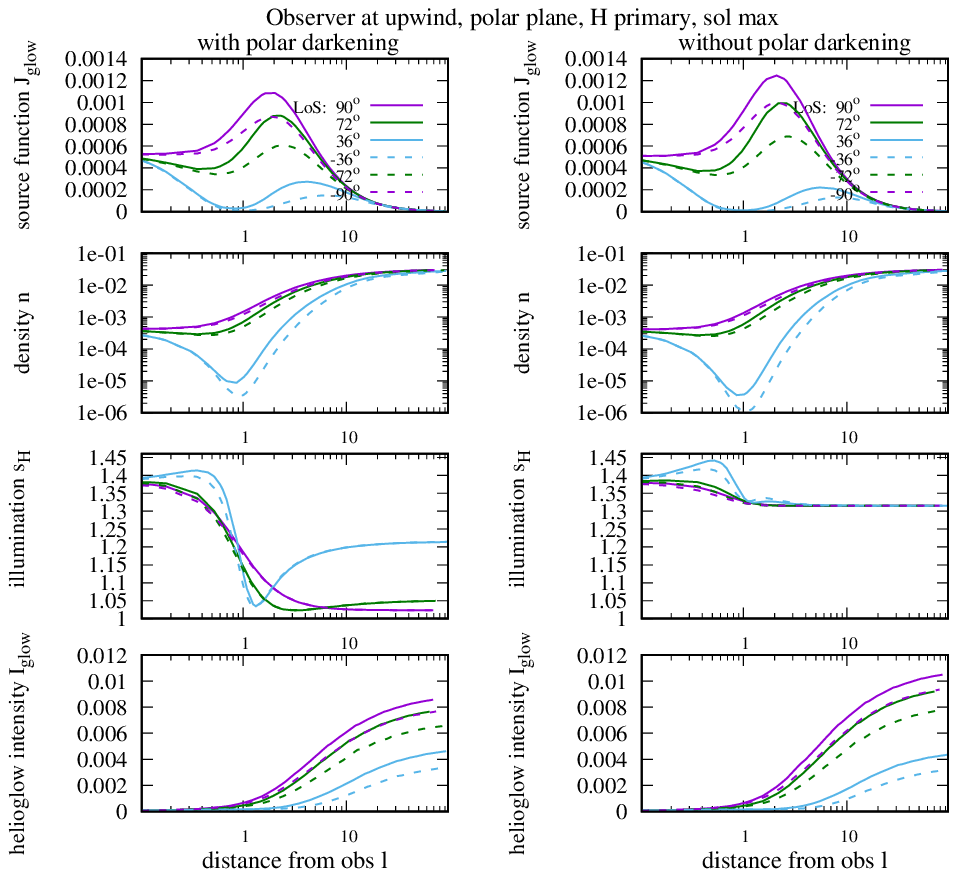}
\caption{Variation of selected elements of the helioglow source function for the location of the observer at 1 au upwind in the ecliptic plane, for solar maximum conditions, the primary population. The left column presents the case with the polar anisotropy of the illumination, the right column that with spherical symmetry of these factors assumed. The lines of sight are within the polar plane (see Figure \ref{fig:gridSelect} and the upper right panel in Figure \ref{fig:densPlane}). The offsets of the lines of sight from the scanning circle center are listed in the upper left panel, positive to the north (solid lines), negative to the south of the Sun (broken lines). The quantities presented include, from top to bottom: the source function $J\glow$, the density $n_\text{H}$, the illumination $s_\text{H}$, and the partial helioglow intensity $I\glow$, shown as a function of the distance along the line of sight $l$.}
\label{fig:upmaxPolespr2D3D}
\end{figure}

\subsection{Effect of the latitudinal anisotropy in the solar EUV output}
\label{sec:illuminationFlattening}
\noindent
The anisotropy of the solar EUV output potentially affects the helioglow in several ways. The first and direct effects is the illumination: whatever the gas distribution in space is, it is illuminated differently at the poles and differently at the solar equator. However, the solar EUV radiation is also responsible for photoionization of ISN gas and for radiation pressure. Both these factors affect the distribution of the gas in space. It can be expected that the latitudinal anisotropies of the solar output at different wavelengths, if not identical, are at least directly proportional. However, their effects on the helioglow need to be carefully studied. Reducing the EUV flux above the solar poles relative to that at the equator will result in reducing the illumination, but also reducing the radiation pressure and the ionization losses, which in turn is expected to increase the gas density in these regions and to compensate for the reduced illumination, at least partly. The effect requires a careful modeling, which we performed in this study.

To investigate the effect of a latitudinal anisotropy of the solar EUV output, we simulated the helioglow either including or excluding this effect. Including the EUV anisotropy modifies the solar illumination of the ISN gas and the latitudinal anisotropy of the ionization rate. According to the adopted model of the H ionization rate, photoionization in the ecliptic plane contributes from $\sim 15\%$ to $\sim 30\%$ to the total ionization rate in 1996 and 2001, respectively \citep[Figure B1 in][]{sokol_etal:20a}. In the absence of the anisotropy of the EUV flux, these ratios at the solar poles are equal to $\sim 30\%$, respectively (i.e., no change between the times of high and low solar activity). The latitudinal modulation of the photoionization rate is identical as in Equation~29 and Figure~\ref{fig:helioLatProfiles} in Paper 1 changes these relations, appropriately reducing the contributions from photoionization at the poles and modifying the density of ISN H in the polar regions. 

Another effect is a change in the illuminating flux. Since the total EUV flux affects the solar spectral flux non-linearly, the reduction in the spectral flux at the poles is by 25\% in 1995 and 20\% in 2001. 

The effects of the EUV anisotropy on the production of the helioglow are illustrated in Figure \ref{fig:upmaxPolespr2D3D}. Clearly, the location along the line of sight of the maximum of the source function $J\glow${} varies very little, but the magnitude of the maximum does change for all line of sight elongations (the top row in Figure \ref{fig:upmaxPolespr2D3D}). The effect on the density along the line of sight (the second row in the figure) in the region of the maximum of $J\glow${} is visible for the lowest-elongation lines, but almost absent in the other ones, in particular for the elongations close to these of the future GLOWS experiment. Most of the effect is due to the change in the illumination itself. This is shown in the third row of Figure \ref{fig:upmaxPolespr2D3D}. As a result, a difference between the total intensities is expected to be similar to that of the anisotropy ratio $a$. More discussion of the anisotropy effect is offered in Section~\ref{sec:SWANcomparison}.

\subsection{Effects of the vantage point location on the observed helioglow}
\label{sec:vantagePointEffect}
\begin{figure}[ht!]
\centering
\includegraphics[width=1.3\textwidth]{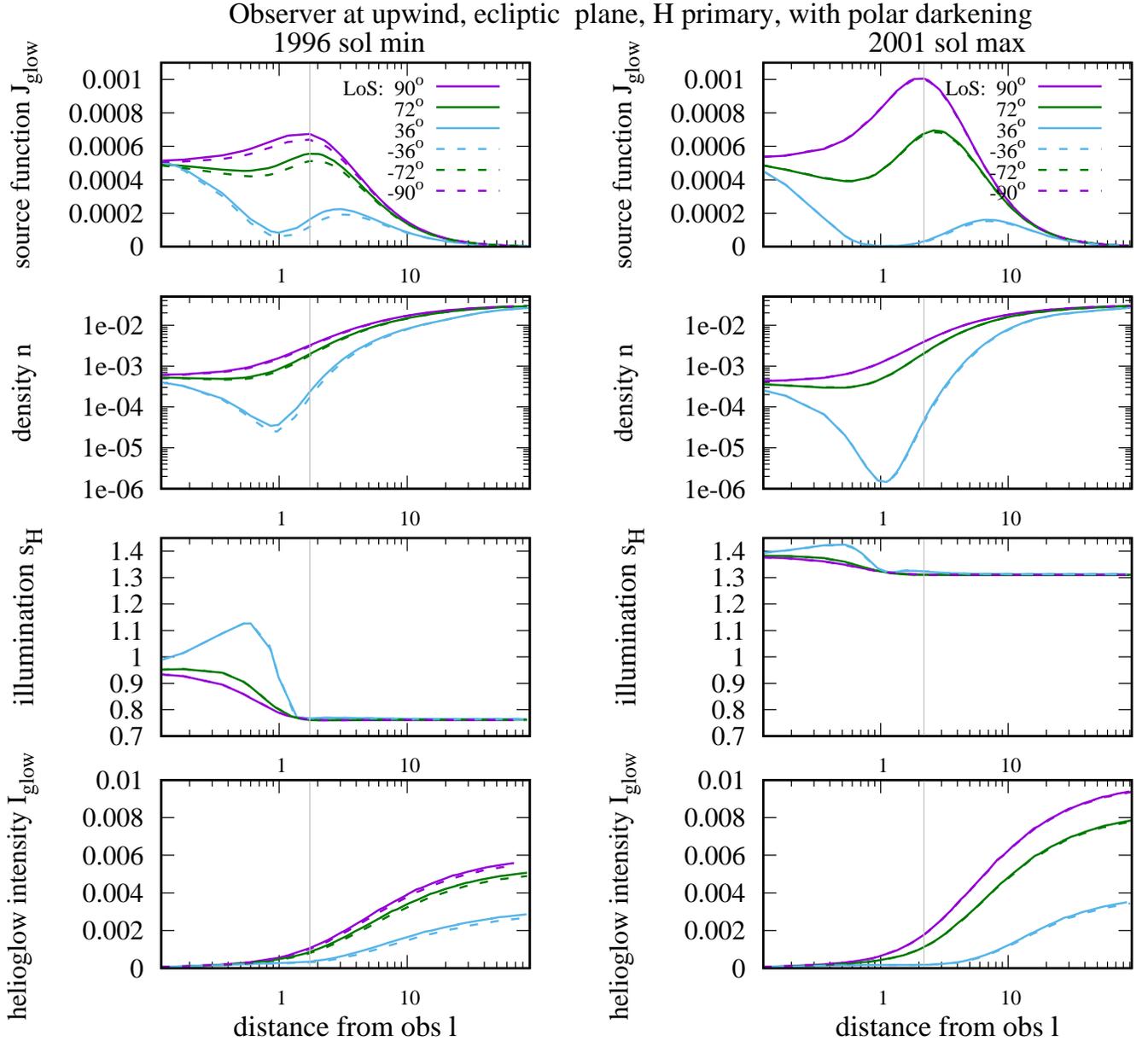}
\caption{Selected elements building up the helioglow: the source function (top row), the gas density (second row), the illumination $s_\text{H}$ proportional to the $\mu$ factor (third row), and the partial helioglow intensity as a function of distance along the line of sight (bottom row), for the solar minimum and maximum conditions (left and right columns, respectively). The observer is at 1 au in the ecliptic plane upwind, the lines of sight are within the ecliptic plane. The elongation of the lines of sight from the Sun are presented in the upper-left panel. The vertical bar marks the location of the maximum emissivity region for the line of sight $-90\degr$.}
\label{fig:upminmaxEquatorialpr}
\end{figure}

\begin{figure}[ht!]
\centering
\includegraphics[width=1.3\textwidth]{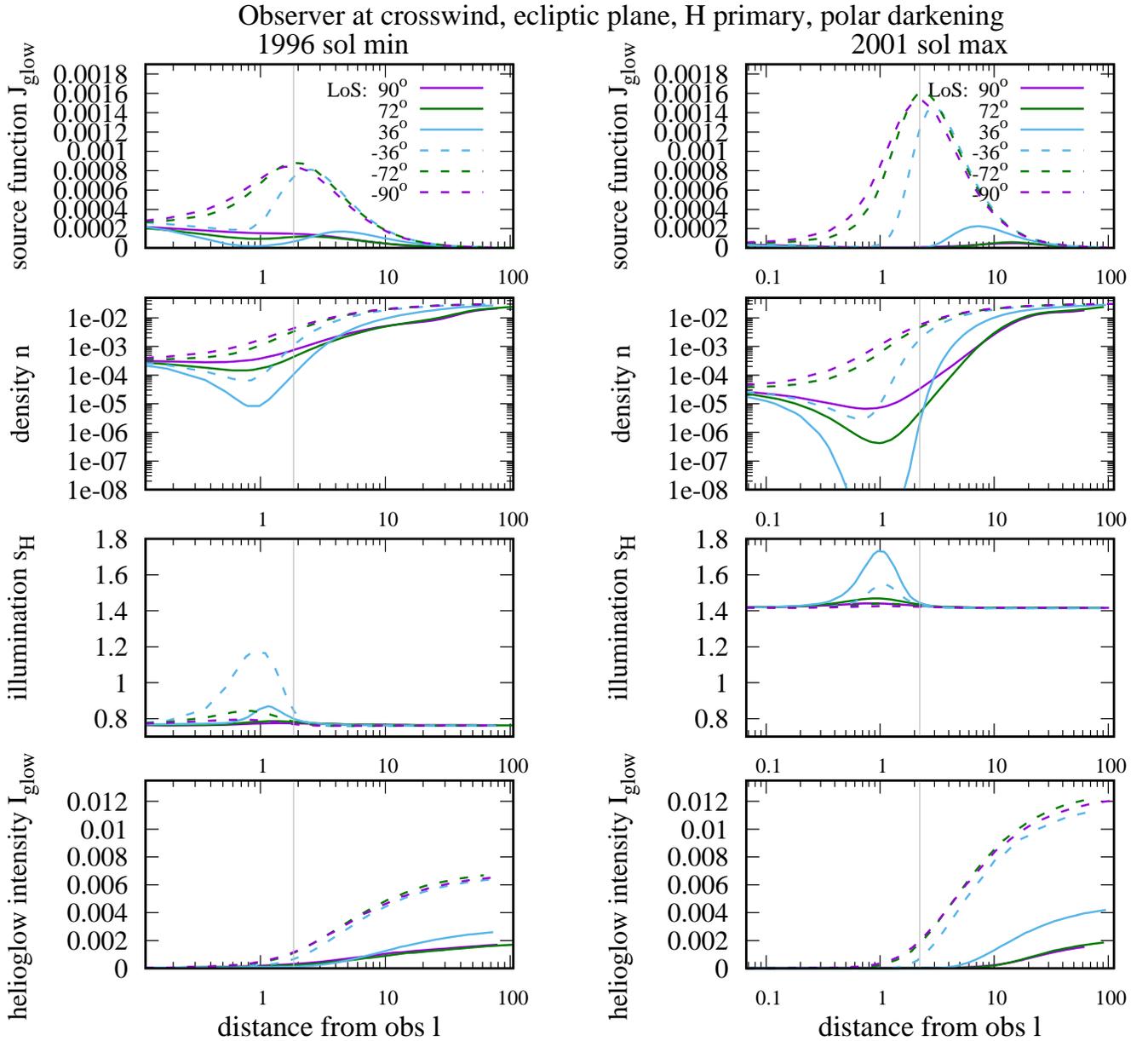}
\caption{Equivalent to Figure \ref{fig:upminmaxEquatorialpr} for the crosswind vantage point. }
\label{fig:cwminmaxEquatorialpr}
\end{figure}

\begin{figure}[ht!]
\centering
\includegraphics[width=1.3\textwidth]{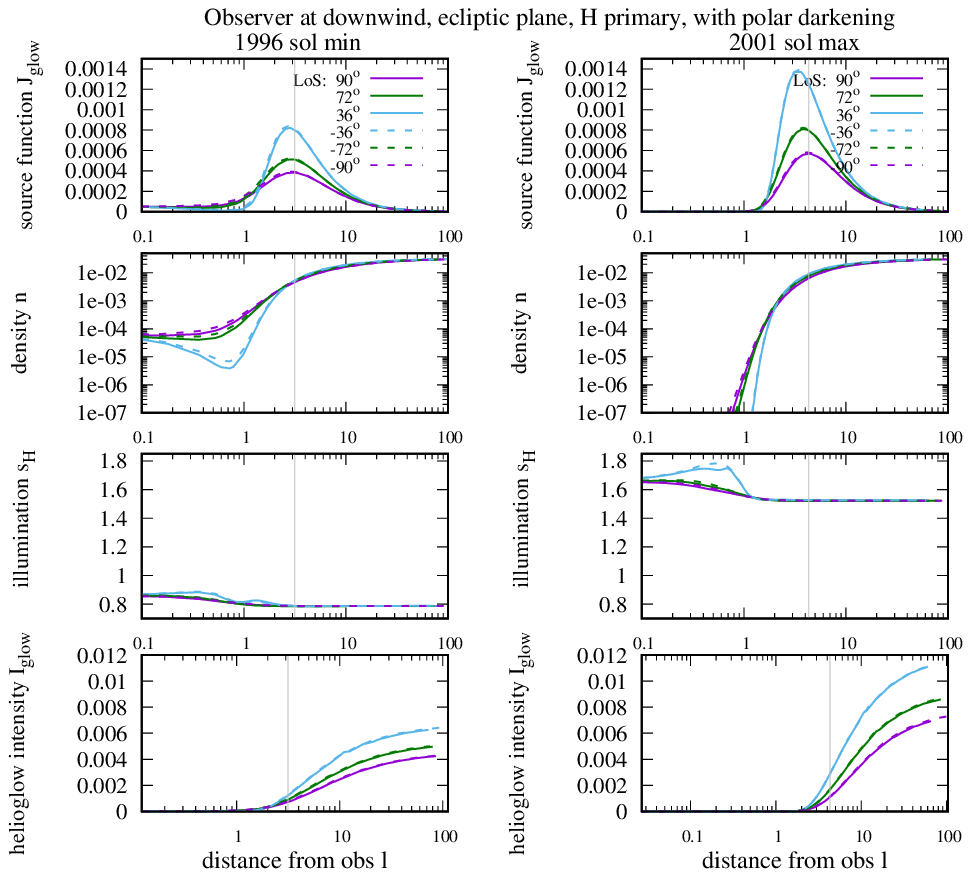}
\caption{Equivalent to Figure \ref{fig:upminmaxEquatorialpr} for the downwind vantage point. }
\label{fig:dnminmaxEquatorialpr}
\end{figure}

\noindent
The formation of the helioglow signal along the line of sight strongly depends on the observer location relative to the inflow direction of ISN gas, i.e., on the vantage point. This is illustrated in Figures \ref{fig:upminmaxEquatorialpr}--\ref{fig:dnminmaxEquatorialpr}, which provide a comparison of the source function, gas density, solar illumination, and the helioglow intensity buildup as a function of distance along the line of sight for the vantage point at the upwind, crosswind, and downwind points at 1 au in the ecliptic plane. Here, we only point out to the most salient aspects. 


First, it is important to realize how the points belonging to the lines of sight for an observer at the upwind, downwind, and crosswind locations are distributed in space. One can think of the lines of sight discussed in the aforementioned figures as a bunch of lines extending from the observer location. Going from the upwind to the crosswind and downwind locations rotates this bunch along the ecliptic with the pivot point at the Sun's center. Within the ecliptic plane, the right-hand and left-hand elongations offer a mirror symmetry for each other for a vantage point either upwind or downwind. One should note that the exact symmetry is present in terms of geometry only, while for spatial distribution of the gas and its illumination, the symmetry is only approximate. The left-hand elongations for the upwind vantage point have their equivalents as the right-hand elongations downwind. The same is true for the two crosswind vantage points at the ecliptic; in the paper we show only one of them. An example of the ecliptic bunch of lines of sight is shown for the upwind observer in Figure \ref{fig:densPlane} (upper left panel).

What is different between the different vantage points in the context of the helioglow signal creation is the density and radial velocity distributions of the ISN gas along the lines of sight, which result in different illuminations and, consequently, differences in the source function. For the upwind vantage point (Figure~\ref{fig:upminmaxEquatorialpr}), the observer is immersed in the relatively dense gas as compared to the downwind vantage point. As we go along the line of sight, the densities increase for all lines of sight with high elongations, as those planned for the GLOWS experiment onboard IMAP. However, for lines of sight with low elongations, the densities initially drop as the distance to the Sun rapidly decreases and start to recover approximately at 1 au from the observer. This behavior is qualitatively similar for the solar minimum and the solar maximum conditions. 

Because of the variation in radial speed, the illumination along these lines of sight is also similar. Generally, for the upwind location, the lines of sight probe the downwind hemisphere, where the densities are lower than upwind (if the same distance from the Sun is considered), and because of the solar forcing the spread in the directions of individual atoms is larger. This results in different radial velocities of the gas for the upwind and downwind locations. 

In result, for the upwind observer location and high elongations, the source function values are relatively high already at the vantage point. The function drops significantly below the vantage-point level at distances beyond ${\sim}4$ au for the solar minimum and ${\sim}7$ au for the solar maximum. This means that the formation of the signal starts immediately from the vantage point. However, for the low elongations, the source function shows a qualitatively different behavior, with a maximum at the observer, a drop to a minimum value at 1--2 au, and a buildup to a second local maximum at several au from the observer. This second maximum corresponds to the maximum emissivity region of the ISN gas. The distance to this region strongly varies with the elongation, especially for low elongations. Since far away from the Sun (i.e., outside of the hydrogen cavity) the gas densities are approximately the same, one can expect that the variability of the helioglow signal is generated somewhere closer. In this context, the behavior of the source function in the proximity of the observer is important, even if most of the helioglow is generated farther. On the other hand, measurements from the upwind vantage point can be considered as ``contaminated'' by the local contribution, if more distant regions are of main interest.

For the crosswind location (Figure \ref{fig:cwminmaxEquatorialpr}), there is a large asymmetry in the signal creation between the two sides of the Sun. At the observer site, the density of the gas is comparable to that at the upwind location during solar minimum conditions but very low for the maximum conditions. A portion of the signal is formed directly close to the observer, as was the case for the upwind location. The distance to the maximum of the source function vary little for the lines of sight running upwind from the Sun, but vary profoundly for those running downwind from the Sun, both as a function of elongation and of the phase of the solar cycle.  

For the downwind location (Figure \ref{fig:dnminmaxEquatorialpr}), the viewing geometry makes a mirror symmetry with respect to the upwind one, but the density distributions along the line of sight are very different: they are very low at the observer, and increase in a significant way only starting from $\sim 1$ au down the lines of sight (measuring the significance by the magnitude of the source function). This is true for both the minimum and the maximum of the solar activity, even though details differ between the two phases. As a result, the entire signal reaching the detector is formed far away from the observer (beyond $\sim 1$  au along the line of sight). Thus, the signal observed from the downwind vantage point (Figure \ref{fig:dnminmaxEquatorialpr}) is created in the crosswind locations in the heliosphere and farther upwind; the locations in the downwind hemisphere do not contribute almost at all. 

We have discussed so far only lines of sight presented in Figures \ref{fig:upminmaxEquatorialpr}--\ref{fig:dnminmaxEquatorialpr}, which are located in the ecliptic plane. Out-of-ecliptic effects are discussed separately further in the paper in Section \ref{sec:PolarVsEcliptic}. In this context it is important to note that for all vantage points, the lines of sight at a given elongation and not in the ecliptic plane traverse very similar latitudes out of the ecliptic and have identical distances to the Sun and the phase scattering function. In other words, lines of sight at identical angles along a Sun-centered scanning radius traverse identical ecliptic latitudes and similar heliolatitudes regardless of the vantage point. This is important for the choice of the scanning circle radius planned for the IMAP/GLOWS experiment.

\subsection{The helioglow variation for various phases of the solar activity}
\label{sec:solActivityVariations}
\noindent
Figures \ref{fig:upminmaxEquatorialpr}--\ref{fig:dnminmaxEquatorialpr} also illustrate the differences in the formation of the helioglow signal between the epochs of low and high solar activity. 

The illumination during the solar maximum is larger by a factor of 1.8-1.9 for all observer locations. This is true basically throughout the lines of sight, with some variations within $\sim 1$ au from the observer. The difference in the illumination seems to be the decisive factor in the magnitude of the helioglow, which is larger for the solar maximum for all lines of sight investigated. This is because while the density features large variations between the lines of sight, solar cycle phases, and vantage points, they are to large extent restricted to a region within a few au from the Sun (a little farther downwind) and a large portion of the helioglow signal originates from relatively remote distances outside the hydrogen cavity. It is instructive to compare the contributions to the signal from the farthest portions of the lines of sight, where they penetrate regions where solar-cycle-related variations of the density practically cease \citep[see Figures 4 (upper row) and 5 in][]{sokol_etal:19b}.

Another aspect of the variation of the helioglow with the solar activity is variation of its heliolatitude distribution. This is illustrated in Figures \ref{fig:h_comparison_1996} and \ref{fig:h_comparison_2001} -- see the middle and bottom panels in the left column. On one hand, the helioglow clearly reacts to the presence or absence of a latitudinal anisotropy in the illumination, but on the other hand to the different heliolatitudinal structures of the solar wind during the high and the low solar activity. The evolution of this structure adopted in the modeling is presented in Figure 6 in \citet{sokol_etal:20a}.

\subsection{Polar vs ecliptic lines of sight}
\label{sec:PolarVsEcliptic}
\noindent
 \begin{figure}[ht!]
\centering
\includegraphics[width=1.3\textwidth]{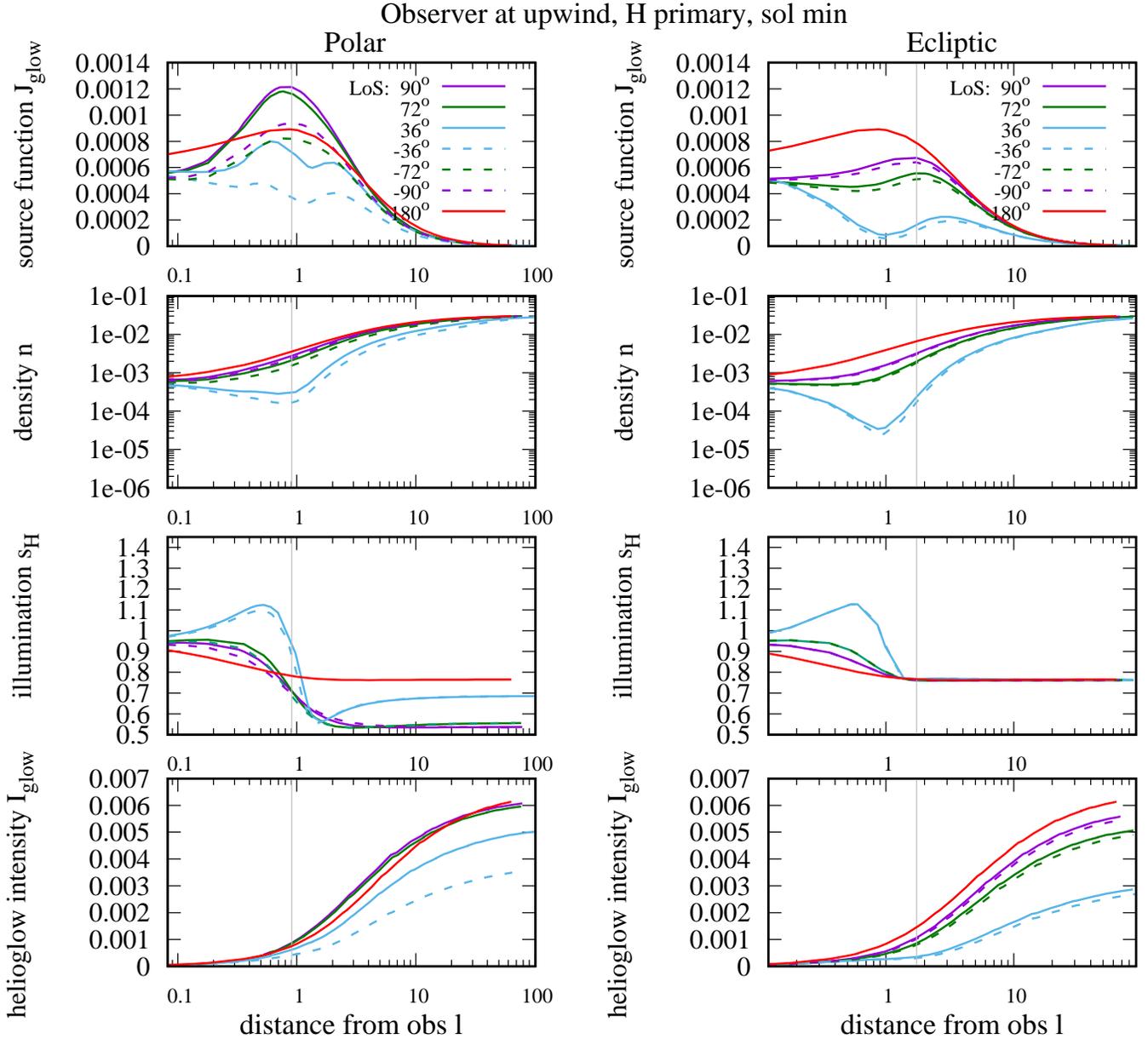}
\caption{Comparison of selected elements making up the helioglow signal for polar and ecliptic lines of sight (cf. Figure \ref{fig:densPlane}, upper panels). The solar minimum conditions, the upwind vantage point. The left column presents polar (north and south) lines of sight  (cf. Figure \ref{fig:densPlane}, the upper right panel), the right column the ecliptic lines of sight. The solar elongations are listed in the upper left panel. From top to bottom: the source function, the density of ISN H, the illumination of ISN H, and the partial intensity, shown along the lines of sight. }
\label{fig:upminPolesEquatorialpr}
\end{figure}
In this section, we point out that lines of sight with carefully selected elongations from the Sun are well suited to investigate the heliolatitudinal effects of the solar radiation and solar wind. We limit the discussion to the upwind viewing geometry; the conclusions for the other geometries are similar. 

The latitudes of the points within the  lines of sight aimed towards low ecliptic latitudes clearly remain low. Thus, the entire information obtained from these lines of sight is obtained from low heliolatitudes. By contrast, the latitude of the points belonging to the lines of sight directed to the north or to the south of the Sun have varying heliolatitudes. This is illustrated in Figure~7 in \citet{sokol_etal:13a}. To maximize a differentiation between the signal from ecliptic and polar lines of sight at identical elongations from the Sun, the elongation angle must be carefully selected. Elongation 90\degr, i.e., directing the line of sight towards celestial poles does not always provide an optimum viewing geometry. For the upwind locations, these lines of sight never cross the solar polar line; the gas contributing to the signal has not undergone the maximum of ionization losses at the polar latitudes. This holds true for almost all vantage points except a region straddling the downwind direction.

Therefore, it seems advantageous for maximizing the effect of polar ionization on the helioglow to aim at an angle less than 90\degr{} from the Sun. A reasonable selection seem to be an elongation of $\sim 70\degr-75\degr$. In this case, the line of sight intersects the solar polar axis close to the region of maximum emissivity. The latitude of the polar lines of sight is presented in the middle right panel in Figure \ref{fig:upminPolespr}. The effects  are illustrated in Figure~\ref{fig:upminPolesEquatorialpr}, which compares the elements comprising the source function along polar and ecliptic lines of sight for the upwind vantage point.  This is true for both the north and the south polar lines of sight. 

It is instructive to compare the signal creation along the lines of sight at the same elongations from the Sun but either maintained in the polar, or in the ecliptic plane (compare the left and right columns in Figure~\ref{fig:upminPolesEquatorialpr}). On one hand, this viewing geometry is very well suited for investigations of the anisotropy of the solar EUV output -- notice the difference in the illumination between the polar and ecliptic lines of sight, which exists even though the radial velocities of the gas are very similar for both geometries/planes. But also the densities differ, which is mostly related to different ionization losses of ISN H, that exist because of the anisotropic charge exchange rate. Therefore, the lines of sight at elongations of $\sim 75\degr${} can be used to investigate the latitudinal structure of the solar wind. Hence, it seems that observations taken along a scanning circle of $\sim 75\degr${} are able to provide sufficient information to establish the latitudinal structure of the solar wind and to study the  solar EUV anisotropy. Such observations are planned for the GLOWS experiment onboard the IMAP spacecraft

\subsection{Two-population vs one-population model}
\label{sec:2popvs1pop}
\begin{figure}[ht!]
\centering
\includegraphics[width=1.3\textwidth]{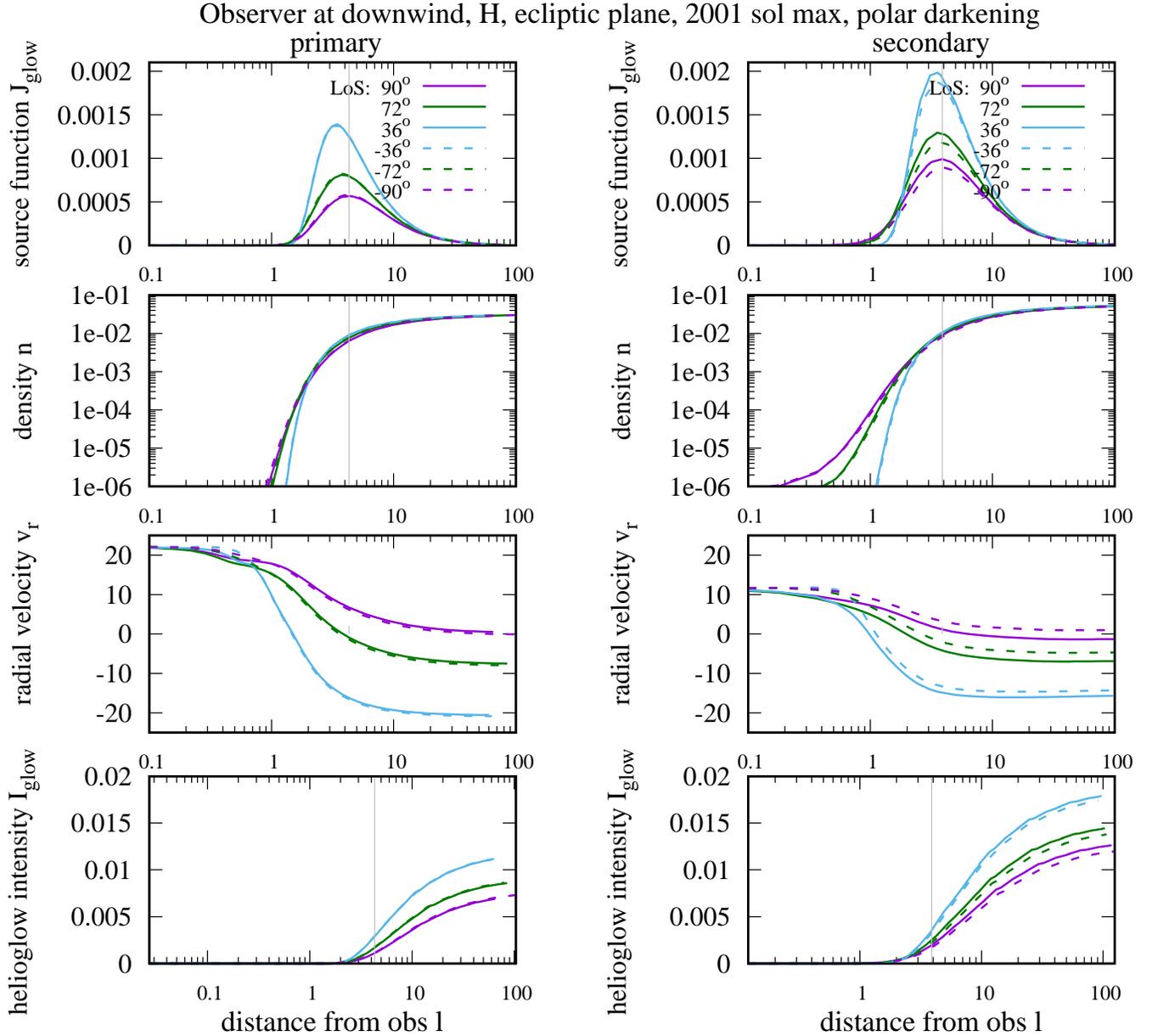}
\caption{Comparison of selected elements building up the contributions to the helioglow signal from the primary (left column) and the secondary components of ISN H (right column). From the top, the panels show the source function, the gas density, the radial velocity, and the partial helioglow intensity at $l$ from the observer. The simulation was performed for the observer at the downwind location during the maximum of the solar activity, the lines of sight are in the ecliptic plane, their elongations from the Sun are indicated in the upper left panel}
\label{fig:dnmaxEquatorialprsc}
\end{figure}

The charge exchange interaction between ISN H atoms and the slowed down and heated interstellar protons in the outer heliosheath results in the creation of a secondary population of ISN H and a strong filtration of the primary population \citep{baranov_malama:93}. 
In this section, we check if the helioglow intensity distribution obtained from the WawHelioGlow model is sensitive to the details of these two populations. We verify a hypothesis that the helioglow intensity observed from 1 au can be simulated using a one-population approach, with the inflow parameters carefully derived from those of the two heliospheric populations of ISN gas. We start, however, with comparing the helioglow signal formation from the primary and the secondary populations (Figure \ref{fig:dnmaxEquatorialprsc}).
For the primary and secondary populations, the adopted parameters were similar to those obtained by \citet{bzowski_etal:08a, bzowski_etal:09a} concerning the densities and the temperatures. The directions were adjusted to better represent the state of the art. The primary population was simulated using the velocity vector determined by \citet{bzowski_etal:15a} for ISN He, and the secondary had a velocity vector identical to that found by \citet{kubiak_etal:16a} for the secondary population of ISN He. While obtained for ISN He, not H, a study by \citet{izmodenov_alexashov:15a} suggests that the angle between the inflow directions of the two populations of ISN H is similar in magnitude to that for He and that these directions should be in the same plane. The magnitudes of the parameters are the following:
primary:  $\lambda = 255.745\degr, \phi = 5.169\degr,  v = 25.784$~\kms, $T =  7443$~K, $n = 0.031$~cm$^{-3}$;
secondary: $\lambda = 251.57\degr, \phi = 11.95\degr, v = 18.744$~\kms, $T = 16300$~K, $n = 0.054189$~cm$^{-3}$. 
The longitudes $\lambda$ and latitudes $\phi$ of the inflow directions are in the J2000 ecliptic coordinates.

The profiles of the densities of the two populations along the corresponding lines of sight are different. The density of the primary at the observer location downwind is almost nil, so almost entire gas population is due to the secondary population. The density gradient of the primary population along the line of sight is much steeper than that of the secondary. However, the behavior of the source functions along $l$ are similar to each other, with the maxima for the two populations located within $\sim 1$ au from each other. Since the density of the secondary population at the termination shock is larger than that of the primary, also the absolute density of the secondary near the region of the maxima of the source function are larger and the magnitude of the source function of the secondary is larger. However, these differences are not large and the contributions from the two populations to the signal can be regarded as comparable.
 
The illuminations of the two populations for this observer location are also comparable because the radial velocity profiles along the selected lines of sight are similar between the two populations. This is specific for the downwind location of the observer because generally, due to the speed difference at the termination shock, the radial speeds differ. The source function becomes significantly larger than 0 at $\sim 1$ au from the observer and at these distances, the radial velocities quickly become similar for a given elongation (except the lowest). As a result, the profiles of $J\glow(l)$ are similar between the two populations, even though the absolute magnitude is larger for the denser secondary population. This analysis suggests that restricting most of the discussion to the primary population is justified.

Now, we compare the sky distribution of the \lya{} helioglow calculated assuming a mixture of the primary and secondary population \citep{baranov_malama:93} with a one-population approach with the parameters carefully selected to best reproduce the superposition of the primary and the secondary flows. The calculations are carried out identically as for the case of the primary and secondary populations.

For the one-population case, the parameters were obtained as weighted averages of the primary and secondary populations: 
$\lambda = 252.5\degr, \phi = 8.9\degr, v = 21.26$~\kms, $T = 12\,860$~K, $n = 0.085$~cm$^{-3}$.  

\begin{figure}[ht!]
\centering
\includegraphics[width=0.8\textwidth]{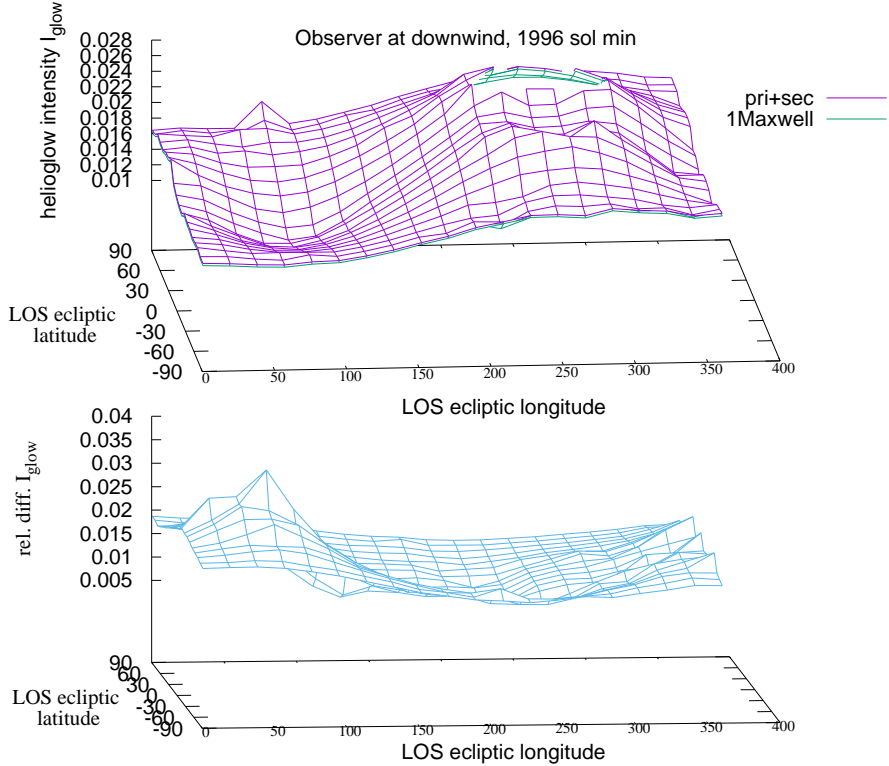}
\caption{A sky map of the intensity of the helioglow for the two-population (purple) and one-population cases (green, upper panel) and the ratio of the differences to the intensity for the two-population case ($1-$1Maxwell/(primary), lower panel). The observer is at the downwind location at 1 au. The units in the upper panel are arbitrary. 
}
\label{fig:figSwanGrida1996a7752aM0064}
\end{figure}
\begin{figure}[ht!]
\centering
\includegraphics[width=0.97\textwidth]{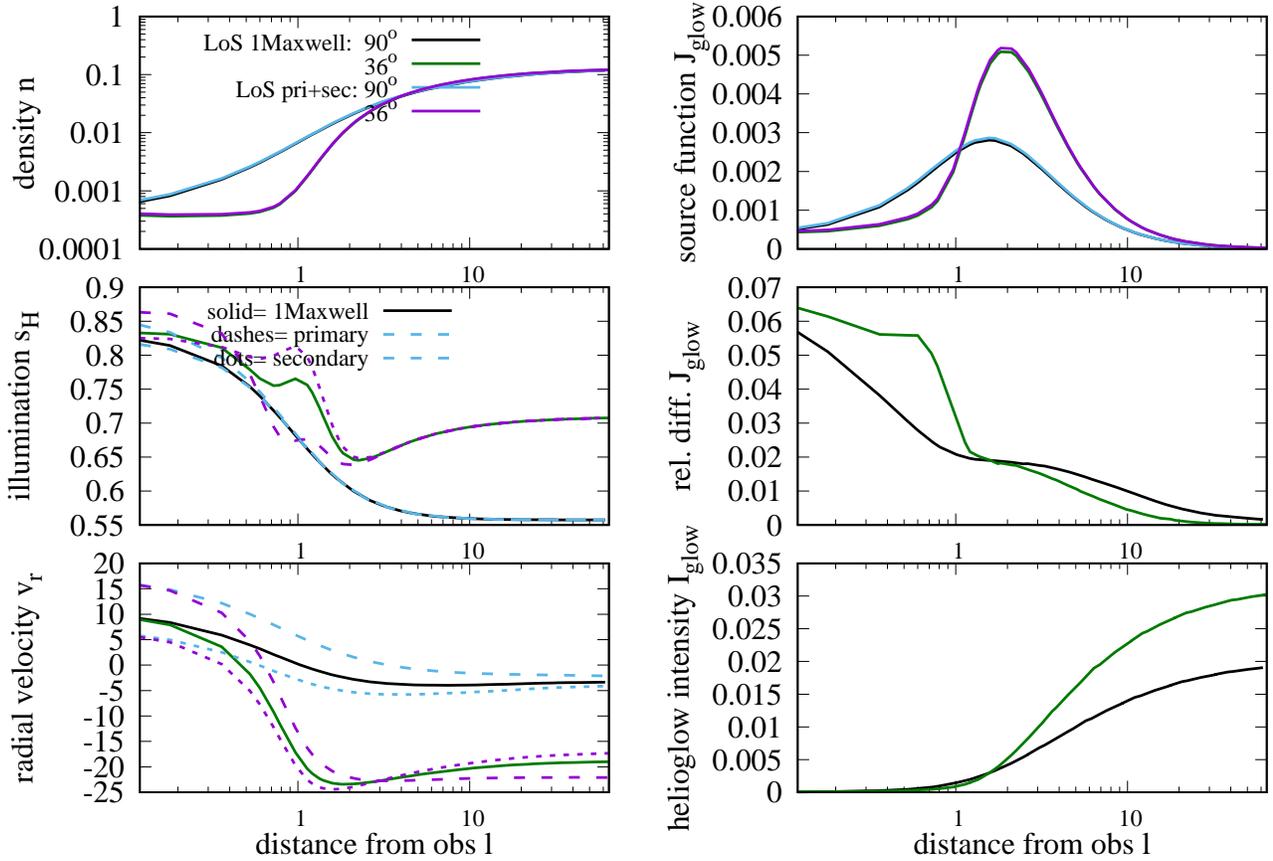}
\caption{Selected elements making up the helioglow signal for the sum of the primary and the secondary populations of ISN H along the lines of sight, compared with their equivalent quantities for the one-population approximation. The observer is at the 1 au downwind position in the ecliptic plane. Two lines of sight are presented: at 36\degr{} and 90\degr{} off the Sun in the polar plane. We show the density (upper left), the illumination (middle left), the mean radial velocity of the gas (lower left), the source function (upper right), the source functions ratio ($1-$1Maxwell/(primary), middle right), and the partial intensity increase with $l$ (lower right). The green and black lines mark the one-population approximation for the elongation angles 36\degr{} and 90\degr, respectively. The purple and blue lines mark the sum of the primary and secondary population for the elongation angles 36\degr{} and 90\degr, respectively. The dashed and dotted lines mark the primary and the secondary populations in the illumination and radial velocity panels.}  
\label{fig:dnPolesSum}
\end{figure}

A sky map of the helioglow for a sum of the primary and the secondary populations and for the one-population approximation is presented in Figure \ref{fig:figSwanGrida1996a7752aM0064}. A comparison of the elements making up the helioglow for the sum of the primary and the secondary populations and for the one-population model for selected lines of sight is shown in Figure~\ref{fig:dnPolesSum}. 
We investigate the differences between the helioglow intensities simulated for the entire sky for these two cases for a vantage point at the downwind location. It is expected that the differences are maximum for this vantage point, since most of the gas contributing to the helioglow signal is already after the closest approach of the atoms to the Sun. Modifications of atom trajectories and the ionization losses are most intense closest to the Sun and the gas density dependence on the input parameters is the strongest.

The differences in the helioglow distributions between the two cases are very small: from 1\% in the upwind region to 2--3\% in the downwind region of the map. We verified that for different vantage points and phases of the solar cycle, the differences are similar. 
The behavior of the factors contributing to the helioglow along selected lines of sight is presented in Figure \ref{fig:dnPolesSum}. The densities agree very well, the source function as well. Larger differences are in the illumination; they result from differences in radial velocities between the populations in question. However, the illumination for the primary and the secondary population average to an effective illumination very similar to that for the one-population case. The differences are smaller than the uncertainty in the model prediction due to the uncertainties in the models of radiation pressure and ionization losses. 

From this insight it seems, then, that analysis of photometric observations of the helioglow, performed from 1 au and aimed at retrieving the ionization structure can be performed using the one-population approach. 

\section{Comparison of modeling results for the WawHelioGlow code with SOHO/SWAN observations}
\label{sec:SWANcomparison}
\noindent
In this section, we present a comparison of full-sky maps obtained from the WawHelioGlow code with selected maps obtained from SOHO/SWAN. We show differences in the helioglow distribution for almost identical locations in the orbit for solar activity minimum and maximum conditions, and demonstrate the effects of including or excluding the latitudinal anisotropy in the solar EUV output discussed in Paper I and in the previous sections.

\begin{figure}[!htbp]
\begin{center}
\includegraphics[width=0.9\textwidth]{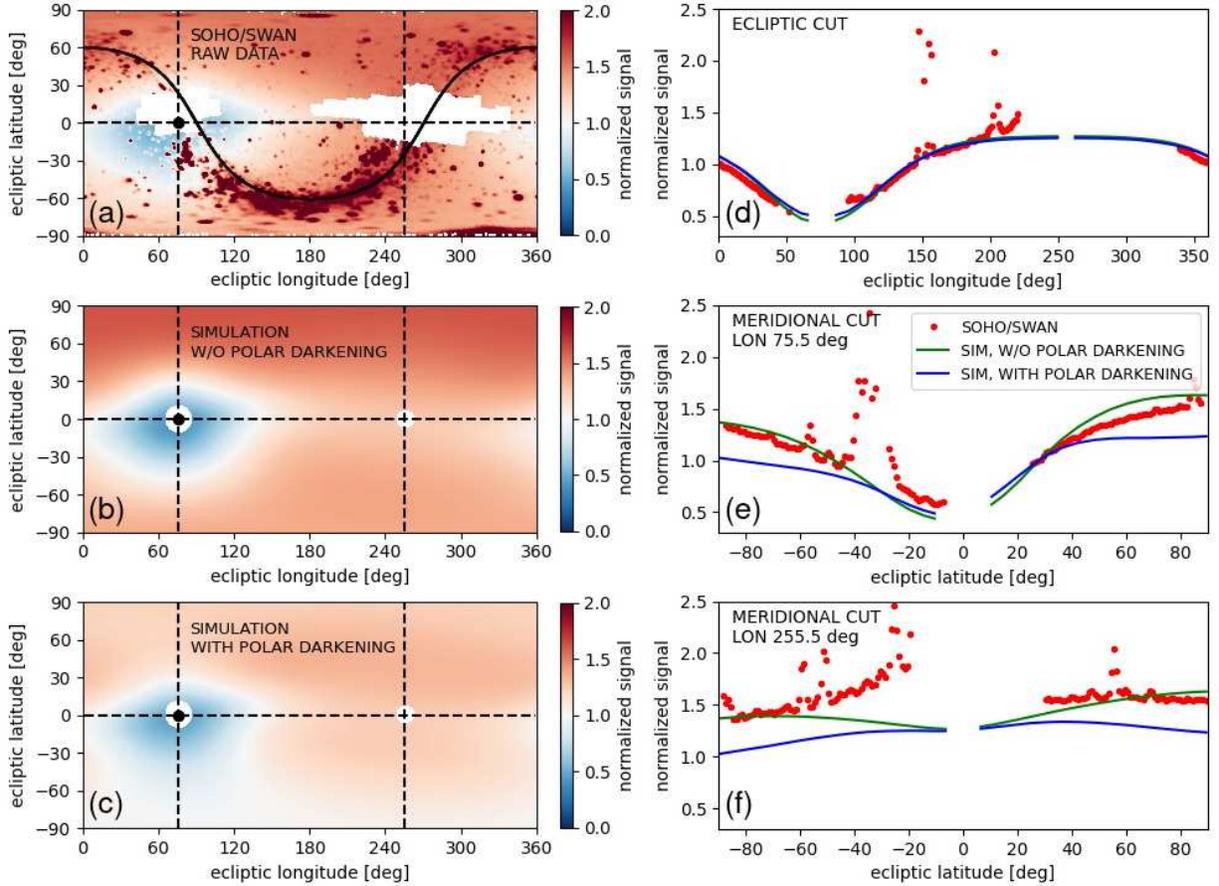}
\caption{Comparison of modeling results from the WawHelioGlow model with SOHO/SWAN observations during minimum of solar activity. The map presented in panel (a)(upper left) shows raw SOHO/SWAN measurements in the ecliptic coordinates, obtained on June 5, 1996, when the satellite was approximately in the upwind location. The white irregular regions indicate the locations of the solar and antisolar masks. The black solid line marks the Galactic equator. The map is normalized (see text). Panel (b)(middle left) shows results of helioglow modeling performed using the WawHelioGlow code with no EUV anisotropy included. Panel (c)(lower left) presents corresponding simulation results with the polar EUV anisotropy included. The white circular regions in panels (b) and (c) are the solar exclusion zone. The Sun location is shown by the black circle, and the dashed lines mark the ecliptic and meridional cuts shown in the right column of plots. In the right column, a comparison of cuts through the maps is shown for (d)(upper) ecliptic, (e)(middle) meridional-downwind and (f)(lower) meridional-upwind cases. In the right-column plots, the red circles represent the SOHO/SWAN measurements (i.e., the sum of the glow and the point-source background, normalized), while the green and blue lines show modeling results for the EUV anisotropy included (blue) or excluded (green).}
\label{fig:h_comparison_1996}
\end{center}
\end{figure}

Figure \ref{fig:h_comparison_1996} presents a comparison of SOHO/SWAN observations and modeling results for June 5, 1996, when the satellite was approximately in the upwind location. This date corresponds to solar minimum conditions. The SWAN/SOHO maps were obtained from the SOHO/SWAN database, where the maps are available in the FITS format \citep{soho_swan:20a,bertaux_etal:95}.

In the raw SOHO/SWAN signal (see Figure \ref{fig:h_comparison_1996}(a), upper left), the \lya{} glow is the slowly varying component, on which signal from point sources is superposed. The point sources are clustered along the galactic equator, as discussed, e.g, by \citet{strumik_etal:20a}. Below the map containing the raw SOHO/SWAN data, results of the helioglow simulation are shown with the anisotropy of the solar EUV output included or excluded (lower left and middle left respectively). In all plots, we use the photon flux density normalized to an average value, taken over the ecliptic plane in the observed or the corresponding simulated map. For the SOHO/SWAN observations, the glow averaging process requires subtracting the point-source background, which was performed using a method proposed by \citet{strumik_etal:20a}. In this method, a phenomenological heliospheric-glow approximator is computed using an approach based on machine learning techniques.

In Figure \ref{fig:h_comparison_1996}(b) (middle left), helioglow simulations without anisotropy in the solar EUV output are presented (i.e., $a = 1$ in Equation~29 in Paper I). Corresponding simulation results with $a = 0.85$ are shown in Figure \ref{fig:h_comparison_1996}(c) (lower left). Visual cross-inspection of the three maps shown in Figure \ref{fig:h_comparison_1996}(a)--(c) confirms general validity of the WawHelioGlow model presented in this paper for description of the SOHO/SWAN observations. A comparison of the mid- and high-latitude regions in the maps reveals that for this particular case, the model without the EUV anisotropy (Figure \ref{fig:h_comparison_1996}(b)) better reproduces the observed modulation of the H glow in the sky as compared with the model with $a = 0.85$, shown in Figure \ref{fig:h_comparison_1996}(c). This conclusion is supported by analysis of ecliptic and meridional cuts through the maps. Figure \ref{fig:h_comparison_1996}(d) (upper right) presents the ecliptic cut through the maps.
The model curves (green and blue lines) show a good fit for the lower envelope of the SOHO/SWAN observations (red circles). These observations are contaminated by the point-source background, which is visible in the cuts as narrow spikes; the model of the helioglow is not expected to fit to these spikes. In the meridional cuts presented in Figure \ref{fig:h_comparison_1996}(e) and (f) (middle right and lower right, respectively), the higher the latitude, the larger the difference between the models, as expected. Clearly, the model without the EUV anisotropy (green line) better reproduces the observations as compared with the model with the anisotropy (blue line). In the meridional cuts, the range of latitudes between $-40\degr$ and 0\degr{} is strongly affected by the presence of unresolved point-sources in the proximity of the galactic equator. Therefore, in this region an apparent discrepancy between the observations and modeling results naturally occurs.

\begin{figure}[!htbp]
\begin{center}
\includegraphics[width=0.9\textwidth]{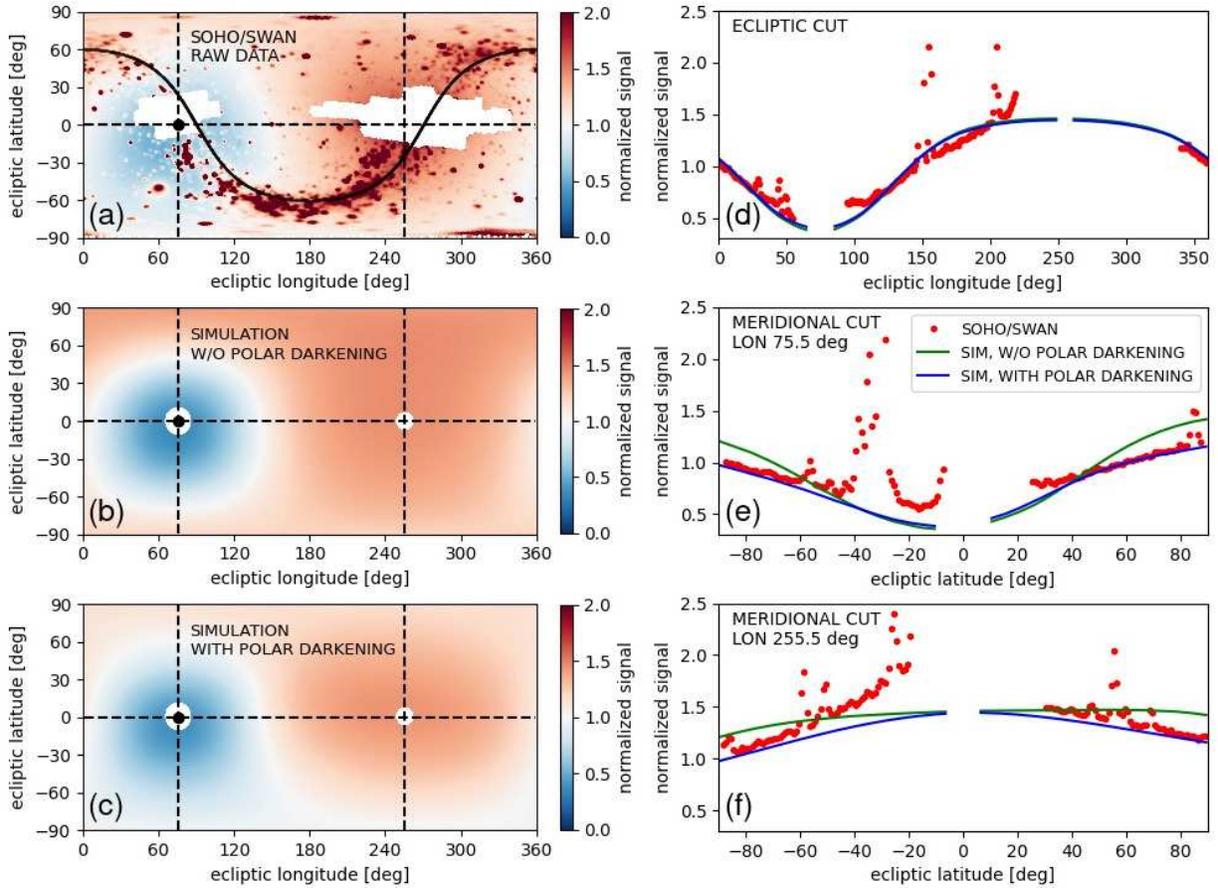}
\caption{Similar to Figure \ref{fig:h_comparison_1996} but for the solar maximum conditions (2001).
The raw SOHO/SWAN measurements presented in panel (a) were obtained on June 5, 2001. The model computations were performed for this date, accordingly.}
\label{fig:h_comparison_2001}
\end{center}
\end{figure}

Figure \ref{fig:h_comparison_2001} shows a comparison of the SOHO/SWAN observations and modeling results for June 5, 2001, which corresponds to solar maximum conditions.
Figures \ref{fig:h_comparison_1996} and \ref{fig:h_comparison_2001}, illustrate the influence of the solar cycle effects on the H glow. The low-emission region in the proximity of the Sun is significantly larger during the solar maximum as compared to the solar minimum conditions. The modulation of the signal changes from an eye-like structure during the solar minimum to a more circular pattern during the solar maximum. Again, the difference between the models with and without darkening is most pronounced in mid- and high-latitude regions. 

This finding is confirmed by analysis of the ecliptic and meridional cuts shown in Figure \ref{fig:h_comparison_2001}(d)-(f) (right row). An important difference with respect to Figure \ref{fig:h_comparison_1996}(d)-(f) is the type of model that fits the SOHO/SWAN observations. For the solar maximum, generally the model with an EUV anisotropy (blue line) better reproduces the observations (red circles). This is in agreement with modeling by \citet{cook_etal:80a, cook_etal:81a}. One should note that in the modeled solar-minimum WawHelioGlow model (see Figure \ref{fig:h_comparison_1996}(b) and (c)) an equatorial attenuation of the signal for the range of longitudes between 180\degr{} and 360\degr{} is visible. Such an effect is not observed for the solar-maximum maps shown in Figure \ref{fig:h_comparison_2001}(b) and (c). Similarly to the solar-minimum case, for the meridional cuts in the range of latitudes between -40\degr{} and 0\degr, the strong discrepancy between the \lya{} glow model and the SOHO/SWAN observations can be attributed to the unresolved point sources in the proximity of the galactic equator.

The insight presented in this section must be regarded as preliminary and incomplete. Without much more extensive analysis, it cannot be determined if the anticorrelation of the solar EUV anisotropy with the phase of the solar cycle is real. Certainly, however, the EUV anisotropy is a frequently neglected effect that needs to be better investigated based on available and future observations, and consequently needs to be implemented in models of the helioglow. 

\section{Buildup of the signal from helium}
\label{sec:SimulationsHe}
\noindent
In this section, we present an analysis of the elements of the source function for the helium glow and we point out important differences between the buildup of the helioglow signal for H and He due to very different spatial distributions of the densities and velocities of ISN H and He close to the Sun \citep[cf Figures 4---6 in][]{sokol_etal:19b} and the very different spectral profiles of the solar \lya{} and He I 58.4 nm lines (cf Figure 3 in Paper I). 

Modeling of the helium glow was carried out to verify the code operation in the absence of radiation pressure and for lines of sight approaching much closer to the Sun than for H, which results in a much more rapid variation with the distance $l$ from the observer. We show that the WawHelioGlow code is capable of taking these effects into account without problem. 

\subsection{Modeling details}
\label{sec:heModelParameters}
\noindent
For the illumination, we used the Gaussian profile (Equation 20 in Paper I) with the background level $s_\text{bkg} = 0.01$. For ionization, we adopted the photoionization, electron-impact, and charge exchange rates from \citet{bzowski_etal:13b} and the latitudinal anisotropy of photoionization as identical to those for ISN H. 
While \citet{sokol_etal:20a} recently updated the ionization rate estimates, especially that of the photoionization rate, details of ionization losses do not significantly modify the discussion of the elements of the helium helioglow source function we present. 

The inflow parameters (the velocity vector and the temperature) were adopted as identical to those for the primary population of ISN H \citep[note that the currently best estimate for these parameters is available from direct-sampling observations of ISN He from IBEX;][]{bzowski_etal:15a, swaczyna_etal:18a}. We neglected the secondary population of ISN He because it contributes very little to the signal, featuring a density of only $\sim 5\%$ of the density of the primary population \citep{kubiak_etal:16a, kubiak_etal:19a}.  

Based on our analysis, for observations at 1 au, it is sufficient to cut off the integration according to Equations 35 and 36 in Paper I at $\sim 20$~au from the observer. Since the size of the helium cavity is much smaller than that of hydrogen \citep[cf Figure 11 in][]{sokol_etal:19b}, in the analysis we used lines of sight much closer to the Sun than those for ISN H, with the elongations equal to 8\degr, 35\degr, and 56\degr. Generally, the distribution of ISN He is very uniform inside the heliosphere, except for two regions: a region inside a few tenths of au around the Sun, where it is depleted, and a cone at the downwind side, where the density is gravitationally enhanced up to a factor of 10. This is in stark contrast to ISN H, where in this region, the density is strongly depleted due to ionization losses and repulsion of atoms by the solar radiation pressure force. 

\subsection{Elements of the source function for the helium glow}
\label{sec:sourceFunctionHe}
\begin{figure}[ht!]
\centering
\includegraphics[width=0.97\textwidth]{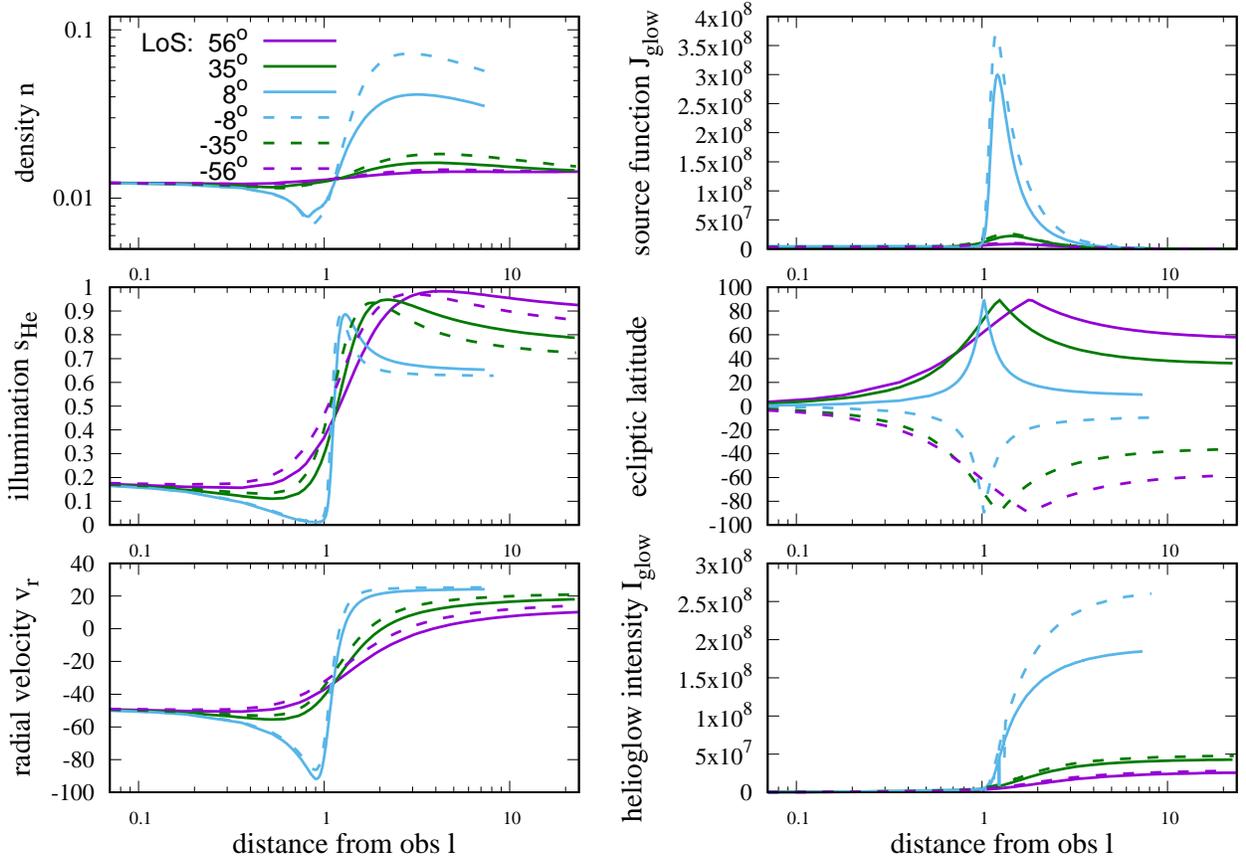}
\caption{Elements of the helium source function for an observer at 1 au upwind for the solar minimum conditions. The lines of sight are in the polar plane, with the solar elongations listed in the upper left panel (positive to the north, negative to the south). The panels present the density $n_\text{He}$ of ISN He at $l$ (upper left), the magnitude of the illumination $s_\text{He}$ (Equation~20 in Paper I; center left), the mean radial velocity of the gas at $l$ (lower left), the magnitude of the source function $J\glow(l)$ (Equation 2 in Paper I; upper right), the latitude of the line of sight $\phi_{\text{He}}$ (center right), and the partial helioglow intensity $I\glow(l)$ (Equation 35 in Paper I; lower right). 
}
\label{fig:HeupminPolespr}
\end{figure}

\noindent
The essential differences between the production of the hydrogen and helium helioglow can be analyzed based on Figure~\ref{fig:HeupminPolespr}, which must be contrasted with Figure \ref{fig:upminPolespr} for H. 
\begin{figure}[ht!]
\centering
\includegraphics[width=0.97\textwidth]{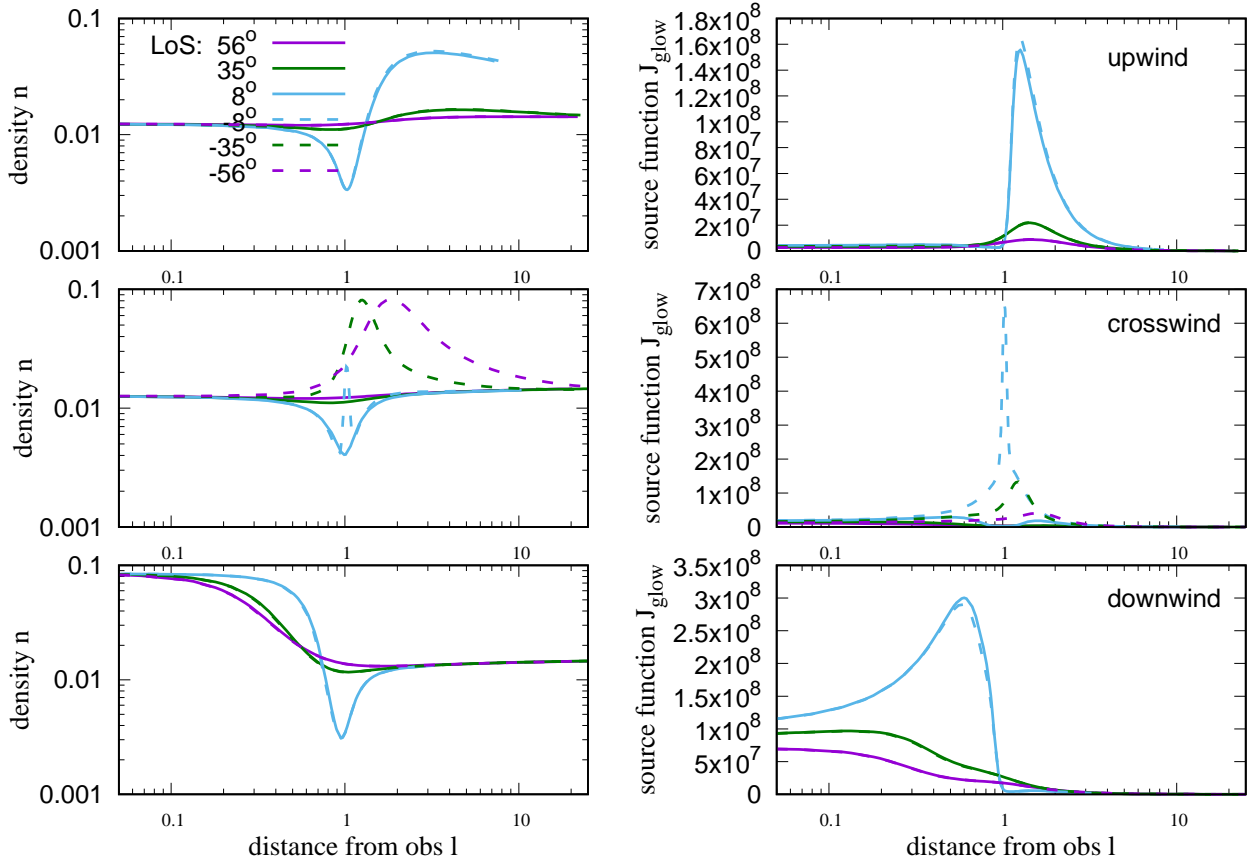}
\caption{A comparison of the densities (left column) and source function magnitudes (right column) along the lines of sight in the ecliptic plane for the observer locations at 1 au in the ecliptic plane upwind (top), crosswind (middle) and downwind (bottom). The elongations of the lines of sight are indicated in the upper left panel. Negative elongation angles mean looking to the left, positive angles to the right of the Sun.}
\label{fig:HeminDensSpr}
\end{figure}

\begin{figure}[ht!]
\centering
\includegraphics[width=0.97\textwidth]{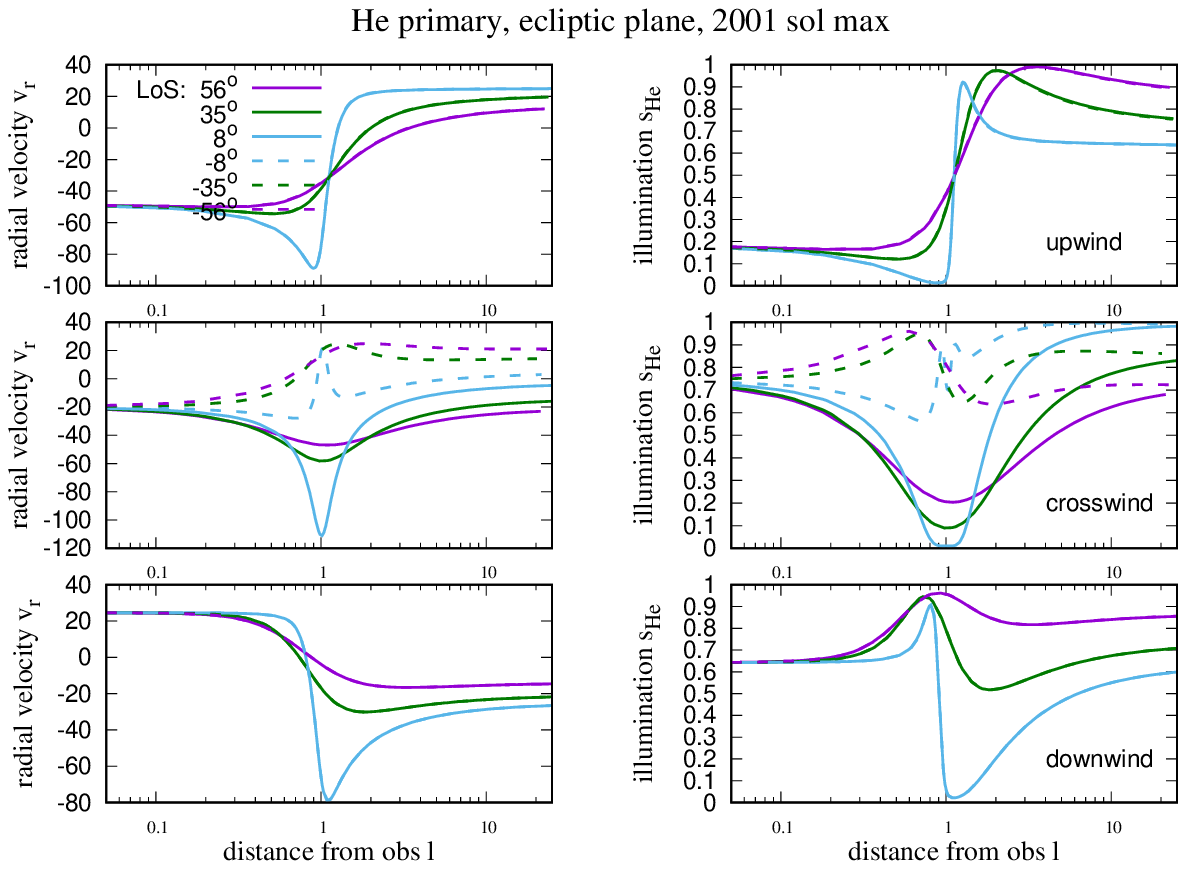}
\caption{For the viewing geometries identical to those presented in Figure~\ref{fig:HeminDensSpr}, the radial velocities (left column) and the illumination (right column) along the lines of sight.}
\label{fig:HeVrlineProfilemaxEquatorpr}
\end{figure}

For large elongations from the Sun (35\degr{} and 56\degr), for observations carried out from the upwind location situated in the ecliptic plane (Figure~\ref{fig:HeupminPolespr}) the distribution of ISN He is almost flat (see the upper left panel). Most of the variation of the source function along the line of sight is, therefore, due to the Doppler dimming effect, the drop of the illumination with the square of solar distance, and the variation of the phase function (Figure~\ref{fig:HeupminPsi}, and Equation 9 in Paper I). Doppler dimming is a well known effect: for the atoms traveling with radial velocities resulting in Doppler shifts comparable to or greater than the width of the illuminating spectral line, the excitation function magnitude is low because only photons from far wings of the solar line are able to excite these atoms. The radial velocity of ISN He, as shown in the lower-left panel, varies from $\sim -50$~\kms{} near the observer up to $\sim 20$~\kms{} for the downwind portion of  the line of sight, which results in a strong modulation of the illumination (the middle-left panel). Note that for the lines of sight with low elongations, the increase in the magnitude of the radial velocity is so large that due to the Doppler dimming only the spectral background of the solar line is able to illuminate the gas. 

The modulation of the excitation function along the line of sight for low elongations is different from that for the high elongations ($\pm 35\degr$ and $\pm 56\degr$) because the low-elongation lines of sight intersect the helium cone, and consequently the density within the line of sight in the downwind region increases very significantly. This, together with the Doppler-related increase in the illumination in a relatively narrow region behind the Sun, results in a sharp spike in the source function. This behavior, however, is only typical for line-of-sight elongations within a few degrees from the Sun. 

For all elongations for the upwind vantage point, the helioglow signal begins to significantly increase only outside a few au from the observer, for the locations in the lines of sight well in the downwind hemisphere. For other locations, i.e., crosswind and downwind, the distribution of the density within the lines of sight is very different, which results in significant differences in the creation of the signal. This is illustrated in Figures~\ref{fig:HeminDensSpr} and \ref{fig:HeVrlineProfilemaxEquatorpr}. At the crosswind location (middle row in the figure), looking at the two sides from the Sun results in the lines of sight going through the upwind region (solid lines) and across the cone (broken lines). The source function is non-negligible starting from the observer location. Then, having passed the upwind/downwind line they drop off relatively quickly, but those going through the cone feature a relatively narrow but strong enhancement, when the maximum of density is traversed. Very different variation in the radial velocity within these lines of sight result in very different illuminations. Anyway, most of the signal is formed within 1--3 au from the observer. 

At the downwind location, the lines of sight located in the ecliptic plane show, not surprisingly, a strong left-right symmetry. Since the observed is now located inside the helium cone, practically the entire signal is formed within 1 au from the observer. The expected north/south asymmetry (not shown) is mostly due to the $\sim 5\degr$ inclination of the flow direction of ISN gas to the ecliptic plane. 

We conclude this section by pointing out that the WawHelioGlow code is able to reproduce the effects expected in the formation of both the hydrogen and the helium backscatter glows. Most interesting effects for He are expected for the lines of sight with relatively low elongations from the Sun. Observing them may be experimentally challenging because of the need to suppress the very strong direct illumination from the Sun in the instrument optics. For H, elongations of 70\degr{} are sufficient, as those planned for the GLOWS experiment on board IMAP. 

\section{Summary and conclusions}
\label{sec:Conclusions}
We presented calculation results from the WawHelioGlow code, which was applied to simulate the helioglow of ISN H and He. We analyzed the elements making up the signal for carefully selected lines of sight and phases of the solar cycle activity. We point out that the source function depends not only on the obvious factors: the gas density and the magnitude of the solar illumination. There are several other factors that vary non-trivially but within similar amplitude ranges, and some of them have been frequently neglected. In the first place, this is the non-flat shape of the solar \lya{} profile, which varies the illumination depending on the radial velocity of the atoms. Another one is the variation of the illumination with heliolatitude. In the paper, we showed how they modify the distribution of the helioglow in the sky for various vantage points and line of sight directions. The sensitivity of the helioglow intensity to all these factors varies, but from those discussed in the papers, we have been unable to identify those that can be safely neglected in all cases.  

This implies that a helioglow model that may successfully be used to interpret helioglow observations should be capable of taking all these effects into account. We show that the WawHelioGlow model is able to do it and that, when used in conjunction with state-of-the-art models of the solar illumination and ionization factors, is able to reproduce full-sky maps of the helioglow, as those obtained from SWAN, with a good fidelity. 

We also demonstrated, based on a limited set of SWAN maps, that simulated maps obtained from WawHelioGlow, in some cases agree better with observations when a simple model of an anisotropy of the solar EUV output is adopted, in addition to the solar wind anisotropy obtained from interplanetary scintillation measurements. In other cases, in a different phase of the solar cycle, the EUV anisotropy in the model seems to be unnecessary. This suggests that the latitudinal anisotropy of the solar EUV output is varying during the cycle of solar activity, which needs to be better investigated, and that photometric observations of the helioglow performed from 1 au, as those available from SWAN and planned for GLOWS can be used to that purpose. We plan to investigate this effect in a future paper.

\begin{acknowledgments}
{\emph{Acknowledgments}}.This study was supported by Polish National Science Center grants 2019/35/B/ST9/01241, 2018/31/D/ST9/02852, and by Polish Ministry for Education and Science under contract MEiN/2021/2/DIR.
\end{acknowledgments}

\bibliographystyle{aasjournal}
\bibliography{helioGlowPaperII}

\end{document}